\RequirePackage{fix-cm}
\documentclass[smallextended]{svjour3}       
\smartqed  
\usepackage{graphicx,natbib,url}
\usepackage{aas_macros}
\begin{document}

\title{Design and Development of Mt.Abu Faint Object Spectrograph and Camera – Pathfinder (MFOSC-P) for PRL 1.2m Mt. Abu Telescope }

\titlerunning{Design and Development of MFOSC-P}        

\author{Mudit K. Srivastava$^{1}$  \and
	Vipin Kumar$^{1, 2}$ \and 
	Vaibhav Dixit$^{1}$ \and
	Ankita Patel$^{1}$ \and
	Mohanlal Jangra$^{1, 3}$ \and 
	A. S. Rajpurohit$^{1}$ \and
	S. N. Mathur$^{1}$        
}
\authorrunning{Srivastava  et al.} 

\institute{
	$^1$  Physical Research Laboratory, Astronomy \& Astrophysics Division, Ahmedabad, India, 380009 \\
	Tel.: +91-79-2631-4401\\
	Fax: +91-79-2631-4900\\
	\email{mudit@prl.res.in} \\
	\and   
	$^2$ Indian Institute of Technology, Gandhinagar,India,382335  \\
	\and
	$^3$ \emph{Present address:} Government Senior Secondary School, Hizrawan Khurd, Fatehabad, India, 125050
}
\date{Received: date / Accepted: date}

\maketitle

\begin{abstract}
	
Mt. Abu Faint Object Spectrograph and Camera – Pathfinder (MFOSC-P) is an imager-spectrograph developed for the Physical Research Laboratory (PRL) 1.2m telescope at Gurushikhar, Mt. Abu, India. MFOSC-P is based on a focal reducer concept and provides seeing limited imaging (with a sampling of 3.3 pixels per arc-second) in Bessell's B, V, R, I and narrow-band H-$\alpha$ filters. The instrument uses three plane reflection gratings, covering the spectral range of 4500-8500$\AA$, with three different resolutions of 500, 1000, and 2000 around their central wavelengths. MFOSC-P was conceived as a pathfinder instrument for a next-generation instrument on the PRL's 2.5m telescope which is coming up at Mt. Abu. The instrument was developed during 2015-2019 and successfully commissioned on the PRL 1.2m telescope in February 2019. The designed performance has been verified with laboratory characterization tests and on-sky commissioning observations. Different science programs covering a range of objects are being executed with MFOSC-P since then, e.g., spectroscopy of M-dwarfs, novae $\&$ symbiotic systems, and detection of H-$\alpha$ emission in star-forming regions. MFOSC-P presents a novel design and cost-effective way to develop a FOSC (Faint Object Spectrograph and Camera) type of instrument on a shorter time-scale of development. The design and development methodology presented here is most suitable in helping the small aperture telescope community develop such a versatile instrument, thereby diversifying the science programs of such observatories.	
	
\keywords{Telescope \and Instrumentation \and Spectrograph \and Camera \and Optical Design}

\end{abstract}

\section{Introduction}
\label{sec-Intro}
The Astronomy and Astrophysics (A$\&$A) division of the Physical Research Laboratory (PRL), Ahmedabad, India (under the aegis of Department of Space, Government of India) operates the Mt. Abu Infrared Observatory (MIRO) at Gurushikhar site (24.653 N, 72.779 E, 1680 m above mean sea level). MIRO consists of a 1.2m Optical-Near Infrared (NIR) telescope equipped with a variety of back-end instruments capable of studying the visible and NIR regime of the spectrum. While the imaging observations in the visible range are being provided by  CCD/EMCCD Cameras, the imaging and spectroscopy in the NIR are being done by the Near Infrared Camera and Spectrometer (NICS) \citep{Anandarao2008}. In addition, a high-resolution echelle spectrograph – PARAS \citep{Chakraborty2014} has also been added to the suite of instruments for exo-planet studies. All these instruments on the 1.2m telescope have been very productive in producing results in diverse areas of astrophysics, e.g., studies of novae and supernovae \citep{Srivastava2016, Joshi2017, Banerjee2018}, Active Galactic Nuclei \citep{Kaur2017}, Comets \citep{Venkataramani2019}, Binary systems \citep{Chaturvedi2016}, detection of exo-planets \citep{Chakraborty2018}, etc. To continue with the tradition of observational research at PRL, the A$\&$A division is establishing another 2.5m optical-NIR telescope at Mt. Abu \citep{Pirnay2018}. The new telescope is expected to see first light in 2021. Four new back-end instruments are being developed for the  2.5m telescope. This includes a Wide-field Camera, a high-resolution spectrograph \citep{Chakraborty2018b}, a NIR imager-spectrograph and an optical imager-spectrograph.
\par
The optical imager-spectrograph for the 2.5m telescope was planned to be developed along the line of the Faint Object Spectrograph and Camera (FOSC) class of instruments. The FOSC series of instruments have been very successful on various telescopes around the world (e.g., EFOSC on ESO 3.6m telescope \citep{Buzzoni1984}, DFOSC on Danish 1.54m Telescope \citep{Andersen1995}, FOCAS on 8.2m Subaru telescope \citep{Kashikawa2000} etc.) as well on several Indian facilities (e.g., IFOSC on IUCAA 2m telescope \citep{Gupta2002}, HFOSC on IIA-HCT Hanle telescope \citep{Prabhu2010},  ADFOSC on ARIES Devasthal 3.6m Telescope \citep{Kumar2016, Omar2017} etc.). A FOSC type instrument has been one of the most sought after general-purpose, versatile instrument on any small or medium aperture telescope, due to its ability to provide imaging and spectroscopy in a single focal reducer based optical chain.  As the requirements of FOSC on the new 2.5m telescope were being planned in late 2014, a similar instrument was also sought for the existing 1.2m telescope. Various science programs of the astronomy group, discussed above, required a facility for low/medium resolution spectroscopy in the optical domain. Thus, as a precursor of a FOSC for the 2.5m telescope, we decided to develop a pathfinder FOSC type instrument for the PRL 1.2m telescope. This pathfinder instrument – named Mt. Abu Faint Object Spectrograph and Camera-Pathfinder (MFOSC-P) - was planned to be a scaled-down version of a typical FOSC, with a relatively smaller budget and development period. MFOSC-P was thus designed and developed with a simple optical design and with a commercially available off-the-shelf detector system. The above choice of these two critical sub-systems led towards substantial design innovation as compared to a typical FOSC type of instrument.
\par 
The design and development routes adopted for MFOSC-P are rather unconventional for a FOSC kind of instrument. Some of the features of MFOSC-P which make it a noteworthy development for the astronomical community; in particular for the users of small aperture telescopes are as follows: a simple optical design, relaxed opto-mechanical tolerances, use of an off-the-shelf detector system and a robust control system based on commercially available components. MFOSC-P has been successfully developed and commissioned on the PRL 1.2m telescope in February 2019 after three and a half years of fabrication and development. The success of MFOSC-P gives a boost to instrumentation for small aperture telescope with limited financial resources. In this paper, we describe the design and development of MFOSC-P, including its optical design, opto-mechanical design, control system development along with laboratory and on-sky characterization tests. We hope that these would be of general interest to the community and enable replication (if so desired) on similar telescopes around the world.

\section{Science requirements and baseline design considerations} 
\label{sec-BaseDesign}

MFOSC-P was conceived to provide low/medium resolution optical spectroscopy and imaging on the PRL 1.2m telescope and to supplement the ongoing observational programs of the A\&A division. Some of the long-term ongoing programs of the division are NIR studies of novae and supernovae using NICS (a hugely successful program), photometric studies of AGNs using CCD cameras, and high-resolution spectroscopy studies of exo-planet/host star using the high-resolution PARAS spectrograph. Thus, the baseline specifications of MFOSC-P were derived from the science needs of such programs. The existing CCD camera used for imaging is equipped with U, B, V, R and I band filters and mounted directly in the telescope focal plane with a field of view (FOV) of $\sim$5.8 $\times$ 5.8 arc-minute${^2}$. NICS (the NIR imager/spectrograph)also has a similar FOV  and uses a 1K$\times$1K Hawaii image sensor for imaging. It provides spectral resolution of $\sim$1000 in spectroscopy mode in $J$, $H$ and $K_s$ wavelength bands from 0.85–2.4 $\mu$m. Therefore, the desired imaging and spectroscopy capabilities of MFOSC-P were chosen not only to complement these existing programs but also to open up new areas of research like optical studies of symbiotic stars. As mentioned in section~\ref{sec-Intro}, the novae and super-novae programs, in particular, have extensively used the NIR spectrometer-NICS. Apart from studies of the novae ejecta, understanding the physical processes and astrochemistry of dust formation, etc.,  NIR spectroscopy has also successfully provided a NIR classification scheme of early novae spectra (especially useful for obscured novae which are faint in the optical but bright in the NIR) \citep{Banerjee2012, Banerjee2018b}. Optical spectroscopy also usefully supplements the early classification of novae spectra into the He/N or Fe II classes \citep{Williams1992}. Identification and evolution of emission lines of various species in novae ejecta, the temporal evolution of continuum, line profile variations seen in various emission lines are useful diagnostics of the evolution of physical conditions in such transient events. These lines cover the complete visible spectrum, e.g., hydrogen Balmer series including H-$\beta$ 4861$\AA$ and H-$\alpha$ 6563$\AA$, low ionization lines of Mg I 5184$\AA$, Na I 5890$\AA$, O I $5577/7773/8446$ $\AA$ along with Fe II multiplets in Fe II class of novae, high excitation lines of N III 4640$\AA$, He II 4686$\AA$, N II 5001$\AA$, N II 5679$\AA$, He I 5876$\AA$, in He/N class of novae, etc. The spectral coverage and resolution of MFOSC-P were thus determined from the requirements of such science cases.
\par 
A typical resolution of $\sim$5$\AA$ over the optical region was considered to be adequate to address the scientific issues discussed above while ensuring a good signal to noise ratio (SNR) over a range of magnitudes (see section~\ref{subsubsec-SkySpec}). This resolution range is also suitable to study potential exo-planet host targets like M dwarfs. A lower resolution of $\sim$10$\AA$ mode was also incorporated for super-novae studies where the emission lines are considerably broad (10,000 $kms^{-1}$ or more). This low-resolution mode was also required to determine the spectral energy distribution (SED) over a broader wavelength range. At the outburst, novae show broad lines with the full width at half maximum (FWHM) of few thousand $kms^{-1}$, which in some cases (e.g., symbiotic novae \citep{Srivastava2015}) decreases with time to few hundred $kms^{-1}$. The evolution of the width and shape of the emission line profiles in symbiotic systems carries significance in understanding the propagation of shock waves into the dense ambient medium surrounding the white dwarf \citep{Bode1985}. To study such a scenario (among other possible ones), a relatively higher resolution mode around H-$\alpha$ was also chosen with a velocity resolution of $\sim100-200$ $kms^{-1}$.
\par 
To minimize the cost of the instrument as well as to keep the fabrication/development period within 2 to 3 years, it was decided to use commercial off-the-shelf solutions for various components and sub-systems wherever possible, e.g., detector system, gratings, filters, calibration optics, motion controllers, etc. Gratings were chosen as the dispersive element instead of grisms, which is a deviation from a customary FOSC design. For similar reasons, a commercially available off-the-shelf CCD camera system, along with its read-out electronics and data acquisition software, was decided to be used in MFOSC-P. However, the use of such a commercial camera system presented a constraint for the instrument's optical system design. The image sensors in these cameras are protected by a glass window in the vacuum enclosures. The glass window is typically kept at a distance of 4-8mm from the sensor. This spacing, as well as the external mechanical mounting arrangements, put a constraint on the minimum distance of the last optical surface of the optical chain from the image sensor. In the optical design of most of the FOSC series of instruments, the last optical element is a corrector lens for field curvature, which is kept close to the image sensor. The distance between these two elements is usually kept small to reduce the aberrations of the imaging system, e.g., field curvature for larger FOVs. It is to be noted that a typical FOSC instrument provides a FOV of $\sim$11 $\times$ 11 arc-minutes${^2}$ on a 2K $\times$ 2K CCD and uses a field corrector lens close to the image sensor\citep{Gupta2002}. We, nevertheless, adopted a FOV of 5.2$\times$5.2 arc-minutes$^{2}$ on a 1K$\times$1K CCD image sensor, with a sampling scale of 3.3 pixels per arc-second (see section~\ref{subsec-OpticalDesign}). The choice of this FOV was consistent with our other existing instrumentation set-up (as mentioned above) and did not require any field corrector lens for aberrations control. It also allowed us to develop a simpler, cost-effective optical design using only spherical singlet and doublet lenses. Further, given the small aperture of the telescope and minimal throughput in the U band, the instrument's optical design was not optimized for the deep blue part or U band of the spectrum.


\begin{table*}
	\centering
	\caption{Design Parameters of MFOSC-P}
	\begin{tabular}{p{5cm}p{5cm}}
		\hline
		\hline
		&          \\
		Parameters &	Values \\
		&          \\
		\hline
		\hline 
		&           \\
		Optimized Wavelength Range &	4500 – 8500 $\AA$ \\		
		Imaging Field of View &	$\sim5.2 \times 5.2$ arc-minute$^2$ \\
		Imaging Pixel Scale	& 3.3 pixels per arc-second \\
		Image Quality Requirement  &	$\sim80\%$ Encircled Energy Diameter is to be within 1.5 pixels \\
		Magnification of the MFOSC-P optical Chain &	0.57  \\
		Camera F/Number	& 7.4 \\
		CCD Detector	& 1K $\times$ 1K ANDOR CCD with 13$\mu$m pixel size \\
		Filters &	Astronomy standard Bessell’s BVRI filters. Narrow band H-$\alpha$ filter \\		Spectral Coverage of Gratings &	$\sim$4500-8500$\AA$ using three different Gratings \\
		Spectral Resolutions $(\lambda/\delta\lambda)$ &	$\sim$2000, 1000 and 500 around 6500$\AA$, 5500$\AA$ and 6000$\AA$ respectively for 1 arc-second ($\sim76\mu m$) slit width \\
		Pupil Diameter & $\sim$33 mm \\
		Grating’s Specifications & Three Plane reflection gratings, named R2000, R1000 and R500 with 500, 300 and 150-line pairs per mm blazed at $\sim$6500$\AA$, 5500$\AA$ and 6000$\AA$ respectively \\ 
		
		\hline		
		\hline
		
	\end{tabular}
	\label{table-para}	
\end{table*}

\par  
Thus, MFOSC-P had been conceptualized to provide grating-based spectroscopy and seeing limited imaging in standard Bessell’s B, V, R, and I filters. A narrow band H-$\alpha$ filter had also been added to the design. The wavelength range for the spectroscopy modes was chosen to be 4500-8500$\AA$, which is sufficient to address the science goals, as described above. Three plane reflection gratings with different groove-densities, viz. 500 line-pairs (lp)/mm, 300 lp/mm, and 150 lp/mm (hereafter R2000, R1000, and R500 grating modes respectively) were selected to cover this wavelength range. R2000 grating mode of MFOSC-P covers the wavelength range of $\sim$6000-7000$\AA$ with a resolution of $\sim3.3\AA$. This corresponds to a velocity resolution of $\sim$150 kms$^{-1}$ around H-$\alpha$. R1000 grating mode has a resolution of $\sim5.0\AA$ over the wavelength range of $\sim$4500-6500$\AA$. This grating can also be centred appropriately so that H-$\beta$ and H-$\alpha$ wavelengths can be covered in a single exposure. R500 grating mode has the lowest resolution ($\sim$12$\AA$) and the largest spectral range of $\sim4500-8500\AA$ to provide SED and spectroscopy of broad lines astrophysical objects. The optical design of MFOSC-P is discussed in the next section. Table~\ref{table-para} gives primary design parameters of MFOSC-P.

\section{Optical and Opto-Mechanical Design}
\label{sec-DesignFull}

\par
\subsection{PRL 1.2m Telescope} 
\label{subsec-PRLTel}
\par 

MFOSC-P has been designed as per optical and mechanical requirements posed by the PRL 1.2m telescope. The telescope is equatorial mounted. It consists of 1.2m primary and 0.3m secondary mirrors with the final f/13 beam at its focal plane \citep{Deshpande1995}. The focal plane of the telescope can be adjusted by motorized movement of the secondary mirror along the optics axis. MFOSC-P is designed to be mounted on the Cassegrain port of the telescope through some additional mounting arrangements. The telescope has a plate scale of 76 $\mu$m per arc-second and aberration-free FOV of $\sim$10.0 arc-minutes of diameter \citep{Banerjee1997}. The weight of the focal plane instrumentation is typically restricted up to 125 kg. The physical space available beneath the primary mirror cell for any instrumentation set-up is $\sim$1.0m diameter $\times$ 0.8m depth. MFOSC-P was thus designed within such constraints of volume, weight, and other mechanical parameters.

\par
\subsection{Optical Design of MFOSC-P} 
\label{subsec-OpticalDesign}
\par

Like many of the FOSC instruments, MFOSC-P optical design is based on the focal reducer concept. Such designs use the collimator-camera optical sub-systems to map the telescope's focal plane onto the CCD detector. The optical chain provides a suitable magnification/de-magnification of the focal plane to achieve a proper sampling of the seeing disc on the CCD. For the PRL 1.2m f/13 telescope, the seeing of 1.0 arc-second corresponds to 76$\mu$m on the telescope focal plane. This seeing disc is to be sampled by at least 3 pixels on the CCD plane. However, the imaging mode sampling scale of 3.3 pixels per arc-second was chosen due to the gratings' anamorphic magnification ($\sim$0.90-0.95)\citep{Schweizer1979}. In this way, a slit width of 1.0 arc-second maps on to $\sim$3 pixels in the spectroscopy modes. As the pixels are 13$\mu$m $\times$ 13$\mu$m in size, the MFOSC-P optical chain has been designed to provide a magnification of $\times$0.57 to achieve these sampling scales.

\begin{table*}
	\centering
	\caption{Parameters of the gratings used in MFOSC-P. All the  
		gratings are
		procured from Richardson Gratings (M/S Newport Corporation.)}
	\begin{tabular}{lccc}
		\hline
		\hline
		&   &   &    \\
		Parameters &   &  Values & \\
		&   &   &    \\
		\hline
		\hline
		Grating Mode     &  R2000   &  R1000 & R500 \\
		\hline
		Catalog No.  &   53-*-396R & 53-*-204R & 53-*-426R \\
		Rulings (line-pairs per mm) &  500    &  300   & 150 \\
		Grating Order    &  1 & 1 & 1\\
		Blaze wavelength$^a$  ($\AA$)    & $\sim$6500   & $\sim$5500  &	$\sim$8000  \\
		Nominal blaze angle$^a$ (degree)  & 11.1 & 4.7 & 3.4 \\
		Angle of incident (degrees)  &  24.7  &  20.1 &  17.9 \\
		(with respect to grating normal)  & & &  \\
		On-axis wavelength$^{b}$ ($\AA$)   & $\sim$6510   & $\sim$5730  & $\sim$6520  \\
		Angle of Dispersion  (degrees)  &  5.3  &  9.9 &  12.1 \\
		(for on-axis wavelengths)   & & &  \\
		Dispersion ($\AA$ per pixel) & $\sim$1.1  & $\sim$1.9   &
		$\sim$3.8  \\
		Spectral range($\AA$)   & $\sim$5950-7100 & $\sim$4700-6650 &
		$\sim$4600-8450\\
		\hline
		\hline
		
	\end{tabular}
	\label{table-gratings}
	\begin{list}{}{}
		\item a: As per manufacturer's (Richardson Gratings, M/S Newport Corporation) grating specification sheets.
    	\item b: On-axis wavelengths for the given grating mode would fall in the centre of the detector in the spectral direction. For these, the sum of angles of incident and angle of dispersion would be 30 degree.
	\end{list}
\end{table*}


\par
\begin{figure*}
	\centering
	\includegraphics[width=0.90\textwidth]{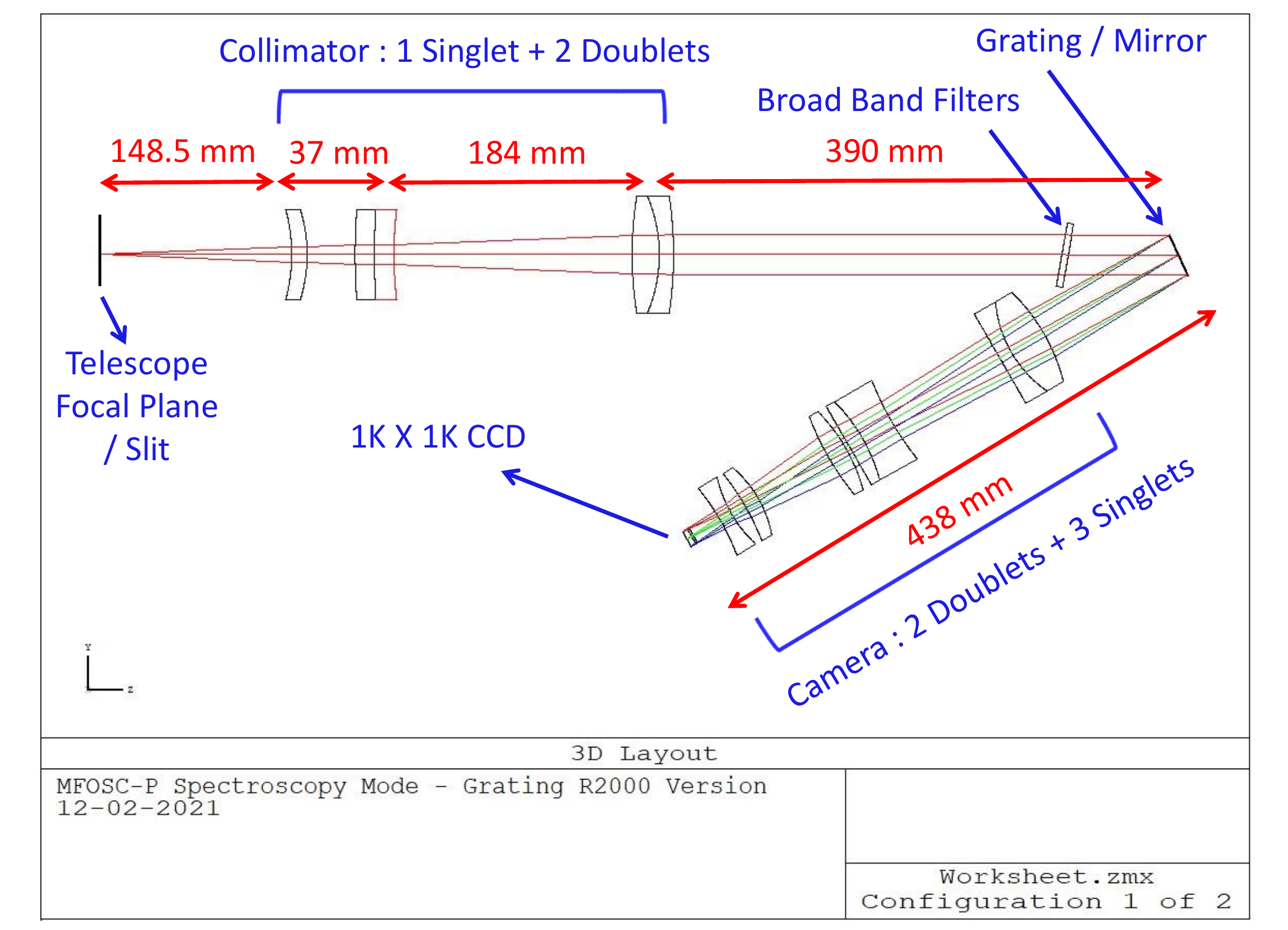}
	\vspace{0.2cm}
	\caption{The optical design of MFOSC-P. An additional fold mirror (not shown in the figure) is also kept immediately after the last lens of collimator to fold the design mechanically. Distances mentioned here are to show the scale of the optical system. Exact values are given in Table~\ref{table-OpticsData}.}
	\label{fig-OpticalDesign}
\end{figure*}

The optical design of MFOSC-P is shown in Figure~\ref{fig-OpticalDesign}. The optical design has been developed in-house using only the spherical lens surfaces. ZEMAX Optical Design Software is used for design purposes. The collimator section is designed using a singlet lens and two cemented doublets lenses to produce a collimated beam. It is kept at a distance of 148.5mm from the focal plane of the telescope. The collimator optics forms the pupil image (diameter $\sim$33.0mm) at a distance of about 385.0mm from its last lens element. As discussed in section~\ref{sec-BaseDesign}, three plane reflection gratings are selected from the catalogue of M/S Newport Cooperation \footnote{https://www.newport.com/b/richardson-gratings; Accessed 2020-03-18}. The important grating parameters, like the blaze wavelengths, efficiencies, wavelength range, and dimensions, are found suitable to be used within MFOSC-P. These gratings are used in the first order to provide the slit-limited resolutions of $\sim$2000, 1000, and 500 at the wavelengths of 6500$\AA$, 5500$\AA$, and 6000$\AA$ respectively. These gratings and a fold mirror for imaging mode are mounted on the purpose-designed turret-like mechanism. This mechanism is rotated by a stepper motor so that the desired grating or mirror can be placed at the pupil position. Optics axes of collimator and camera system are designed to be kept at an angle of 30 degrees to reduce the effects of grating's anamorphic magnification\citep{Schweizer1979}. Thus, for the on-axis wavelengths of the gratings modes (in the centre of the detector on the spectral axis), the sum of angle of the incident ($\alpha$) and angle of dispersion ($\beta$) is fixed at 30 degrees. Various parameters of gratings used in MFOSC-P are given in Table~\ref{table-gratings}.

\begin{figure}
	\centering
	\includegraphics[width=0.99\textwidth]{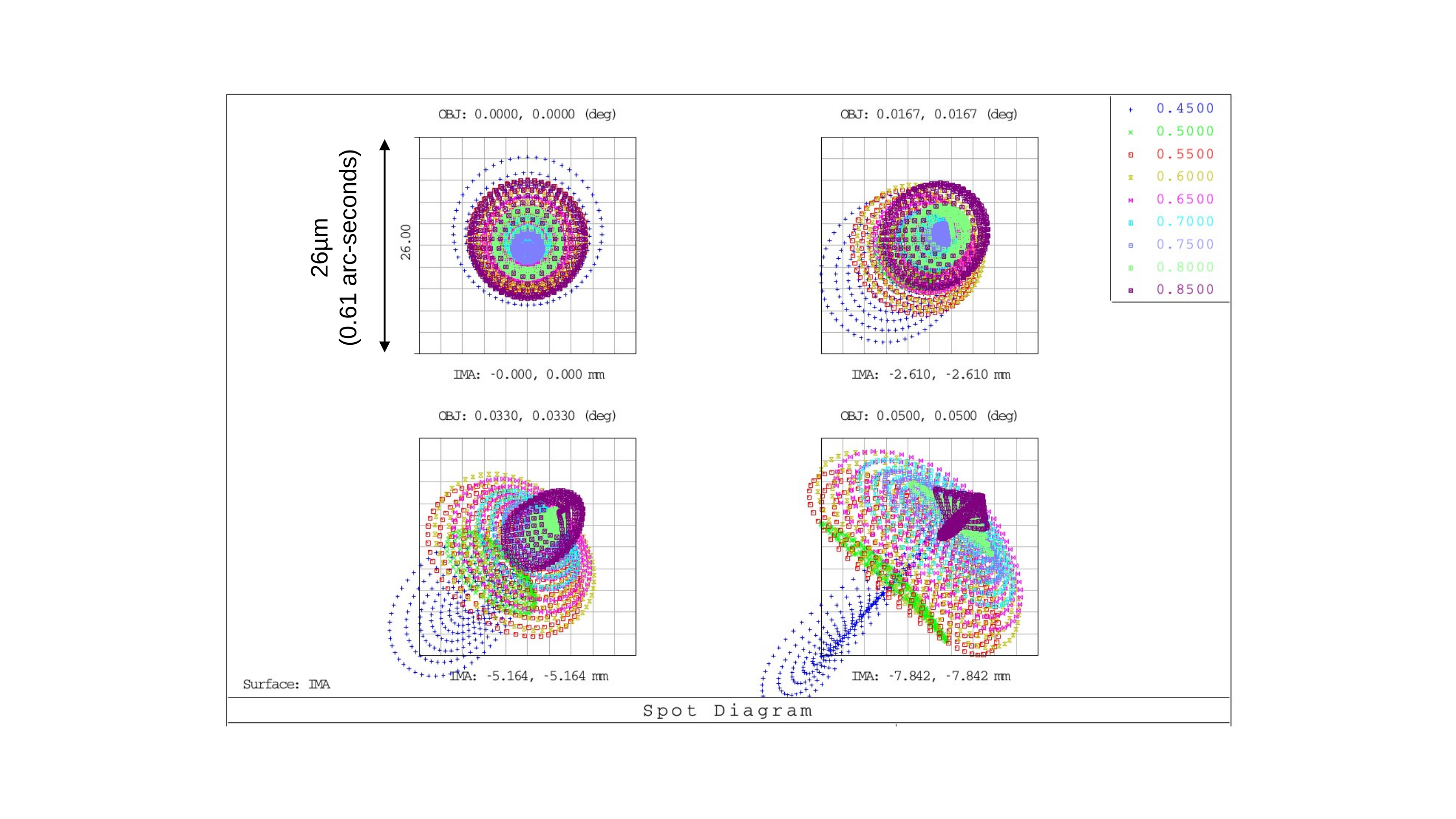}
	\vspace{0.2cm}
	\caption{Poly-chromatic imaging mode spot diagram of MFOSC-P optics at the detector plane, for the design wavelength range 4500-8500$\AA$. Four radial field points are chosen as top-left for on-axis (0,0) arc-minutes field, top-right for (1,1) arc-minutes, bottom-left for (2,2) arc-minutes and bottom-right for (3,3) arc-minutes field coordinates. Box size is 26 $\mu$m (2 pixels) a side. The CCD detector has pixel size of 13$\mu$m a side in size and spans $\sim 5.2\times5.2$ arc-minute$^2$. The details of RMS spot diameters for various field points are given in Table-\ref{table-IQ}.}
	\label{fig-SpotDiagram}
\end{figure}

\begin{figure}
	\centering
	\includegraphics[width=0.99\textwidth]{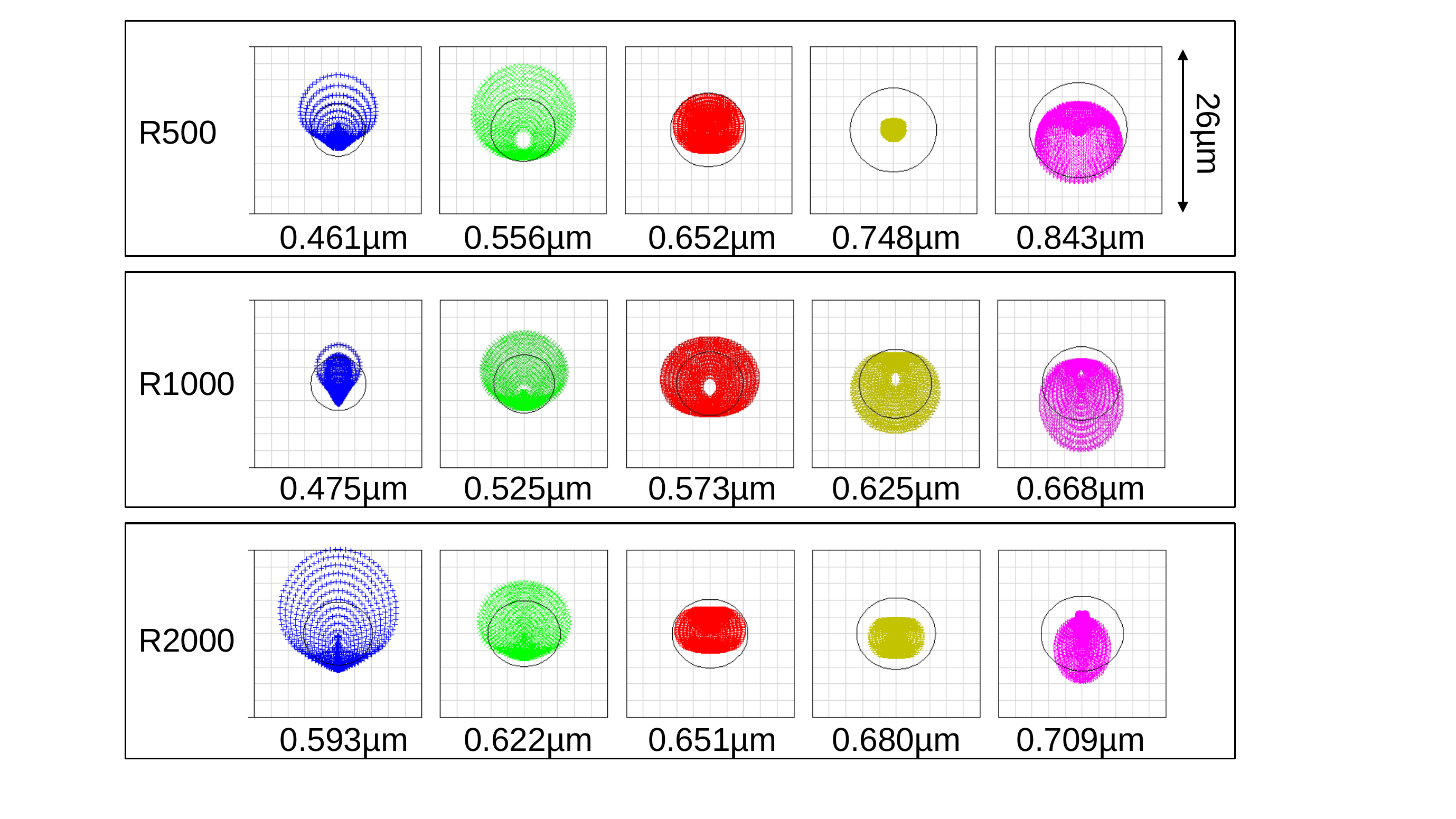}
	\vspace{0.2cm}
	\caption{Matrix spot diagram for R500, R1000 and R2000 grating modes of MFOSC-P. Box size is 26 $\mu$m (2 pixels) a side. The details of RMS spot diameters for all the spectroscopy modes are given in Table-\ref{table-IQ}.}
	\label{fig-MatrixDiagram-R500}
\end{figure}

\begin{table}
	\centering
	\caption{Lens Parameters for the MFOSC-P Optical Chain: This table contains ZEMAX Optical Design Data}
	\begin{tabular}{p{3.0cm}lcccp{2.3cm}}
		\hline
		\hline	
		Element &  Surface & Glass & Catalog & Radius of Curvature & Thickness  \\
		&          &       &         &     (mm)            &   (mm)   \\
		&          &       &         &                     &   (Distance to next Surface) \\
		
		\hline    
		&          &       &         &                     &   \\    
		Collimator-Singlet-1 & First  & N-PK51   & SCHOTT & -135.0 (CC)$^{a}$      & 12.0    \\
		Collimator-Singlet-1 & Second &          &        & -116.0 (CX)$^{a}$      & 37.0    \\
		Collimator-Doublet-1 & First  & S-FTM16  & OHARA  &  478.0 (CX)      & 15.0    \\
		Collimator-Doublet-1 & Second & S-NSL36  & OHARA  &  700.0 (CC-CX)  & 15.0    \\
		Collimator-Doublet-1 & Third  &          &        &  290.0 (CC)      & 184.0   \\
		Collimator-Doublet-2 & First  & S-FPL53  & OHARA  &  286.0 (CX)      & 21.0    \\
		Collimator-Doublet-2 & Second & S-NSL36  & OHARA  & -115.6 (CX-CC)& 11.0    \\
		Collimator-Doublet-2 & Third  &          &        & -307.0 (CX)      & 510.0$^{b}$   \\
		Camera-Doublet-1     & First  & S-FPM3   & OHARA  &  147.0 (CX)      & 29.0    \\
		Camera-Doublet-1     & Second & S-BSM28  & OHARA  & -93.3 (CX-CC) & 10.0    \\
		Camera-Doublet-1     & Third  &          &        &  Inf (Plano)   & 96.8  \\
		Camera-Doublet-2     & First  & N-KZFS4  & SCHOTT & -237.0 (CC)      & 26.0    \\
		Camera-Doublet-2     & Second & K-GFK70  & SUMITA &  135.0 (CC-CX)  & 12.0    \\
		Camera-Doublet-2     & Third  &          &        &  Inf (Plano)   & 3.0     \\
		Camera-Singlet-1     & First  & N-KF9    & SCHOTT &  136.8 (CX)    & 15.0    \\
		Camera-Singlet-1     & Second &          &        & -218.8 (CX)    & 63.0    \\
		Camera-Singlet-2     & First  & N-LASF44 & SCHOTT &  90.4 (CX)     & 11.7  \\
		Camera-Singlet-2     & Second &          &        &  205.3 (CC)    & 12.1  \\
		Camera-Singlet-3     & First  & S-LAH60  & OHARA  & -102.0 (CC)      & 10.0    \\
		Camera-Singlet-3     & Second &          &        &  185.0 (CC)      & 29.65$^{c}$   \\
		\hline
		\hline
		\vspace{-0.2cm}
	\end{tabular}
	\label{table-OpticsData}
	\begin{list}{}{}
		\item a : CC : Concave Surface. CX : Convex Surface. The intermediate surface of a doublet lens is described as CC-CX or CX-CC.
		\item b : The 510.0mm (= 300.0+5.0+85.0+120.0 mm) is the distance between last surface of the collimator lens to first surface of camera lens. 300.0mm is the distance between collimtor to 5.0mm thick imaging filter, 85.0mm is from filter to grating and 120.0mm is from grating to camera lens surface.
		\item c : The last distance from camera lens surface to the image sensor (29.65mm = 24.0+1.5+4.15 mm) is the distance from lens to the protective glass window (24.0mm), thickness of glass window (1.5mm) and from glass window to the image sensor (4.15mm).
	\end{list}
\end{table}


\par
A filter wheel, with Bessell’s B, V, R, I, and H-$\alpha$ filters, is placed before the gratings in parallel beam space. The camera optics is placed at a distance of 120mm from the pupil position after folding the beam. The camera optics is designed with two doublet and three singlet lenses to produce an f/7.4 beam. The image is finally recorded on 1024 X 1024 pixels CCD image sensor with 13 $\mu$m pixel size. The detector system is an off-the-shelf product (Model No. iKon-M934) from M/S Andor Technology Ltd\footnote{https://andor.oxinst.com/products/ikon-xl-and-ikon-large-ccd-series/ikon-m-934; Accessed 2020-03-18}. It consists of a back-illuminated deep depletion CCD and can be cooled to -80 degree C by using Peltier cooling. The read-out noise is 3.3 electrons/pixel at the read-out rate of 50 kHz. The distance between the CCD sensor and the last optical surface of camera optics is optimized to be 29 mm for the reasons discussed in section~\ref{sec-BaseDesign}. Off-the-shelf slits from M/S Thorlabs and M/S Edmund Optics are used at the instrument's object plane. Two slits of 75 $\mu$m and 100$\mu$m width are used in MFOSC-P for varying seeing conditions. The slits are 3mm in length, projecting on to 132 pixels at the CCD. Though the slits do not cover the entire length of the CCD, they were found suitable for the spectroscopy of Galactic point sources and extra-galactic bright super-novae. Given the moderate aperture of the telescope, spatially resolved spectroscopy of extended diffuse sources are usually not done with the telescope. The telescope is also not equipped with a Cassegrain rotator to re-orient the slit. Thus, this off-the-shelf readily available solution for the slits was adopted. Table-\ref{table-para} gives the baseline design parameters of MFOSC-P. The details of optical design parameters, lens specifications, glass types, etc. of the MFOSC-P optical chain are given in Table~\ref{table-OpticsData}.

\par
\subsection{Image Quality and Tolerances} 
\label{subsec-IQandTol}
\par

\begin{figure}
	\centering
	\includegraphics[width=0.85\textwidth]{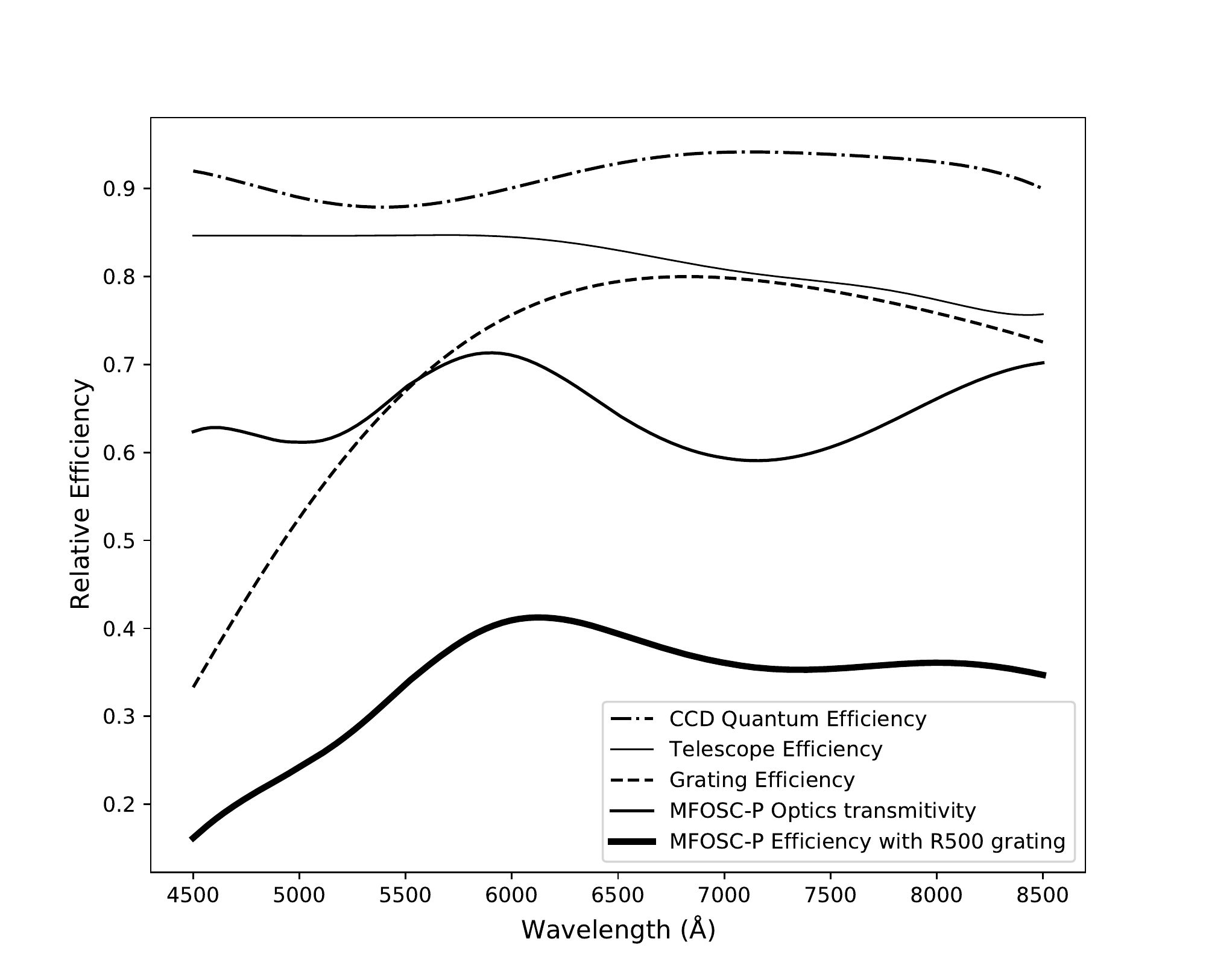}
	\vspace{0.2cm}
	\caption{The throughput variation (solid thick dark curve) of MFOSC-P is shown for R500 grating mode. Various contributors are shown in thin curves. Grating response is for R500 grating. The other modes of spectroscopy have the similar throughput profiles.}
	\label{fig-Throughput}
\end{figure}

MFOSC-P optics image quality requirement was to have the root-mean-square (RMS) spot diameter within one-third of the seeing profile. For the desired sampling of 3.3 pixels per arc-second and seeing of 1.0 arc-second, this translates into an RMS spot diameter requirement of 1.1 pixels. The optical system was thus attempted to design and optimize for RMS spot diameter around $\sim$1.1 pixels, including various components and system tolerances, for all imaging and spectroscopy modes. It must be mentioned that the final PSF on the telescope focal plane also includes the contributions from the telescope optics, its opto-mechanical arrangement, tracking system etc. These extra factors further degrade the FWHM of the PSF on the telescope focal plane (see section~\ref{subsubsec-SkyPhot}). Thus, the RMS spot diameter requirement of $\sim$1.1 pixels truly corresponds to the telescope's seeing limited performance. We, nevertheless, choose this as the design criteria for the pathfinder instrument as the newer upcoming 2.5m telescope would provide seeing limited performance, and the next FOSC instrument for the new telescope would be designed with similar criteria. The optics performance has been optimized for 8.0 arc-minute diameter FOV in imaging mode. The wavelength range of 4500-8500$\AA$ is chosen for the imaging and spectroscopy mode. The polychromatic spot diagram for imaging mode is shown in Figure~\ref{fig-SpotDiagram}. Figure~\ref{fig-MatrixDiagram-R500} shows the matrix spot diagram for the spectroscopy modes. The other two spectroscopy modes have similar ray footprints on the detector plane. Table~\ref{table-IQ} describes RMS diameters for all the modes of imaging and spectroscopy. The throughput of the instrument is estimated $\sim30\%$ for spectroscopy modes. A throughput curve is shown in Figure~\ref{fig-Throughput} for the instrument in R500 spectroscopy mode.

\par 

MFOSC-P optics has been designed to have loose tolerance values considering the typical manufacturing capabilities of the optics manufacturers. Industry-standard manufacturing errors of $\pm25\mu$m /3 fringes on the radii of curvatures and $\pm50\mu$m on the lens elements' thickness were considered for the tolerance analysis. Two components of achromatic doublets were required to be centred with accuracy better than 1.0 arc-minute.  All the lenses were coated with broadband anti-reflection coating with less than $1.5\%$ reflectance over the wavelength range of 4500-8500$\AA$. Several other optics specifications were chosen as per the requirement of scientific imaging applications, e.g., power/irregularity of $4/1$, the surface quality of $\lambda/4$ per inch, Scratch/Dig of 40/20, etc. In addition to component level errors in the manufacturing, tolerances of $\pm100\mu$m and $\pm$0.1 degrees were applied to the positioning accuracy and tilt of the individual optics components, respectively, in their opto-mechanical arrangements. These moderate values of tolerances for opto-mechanical imaging system development are significant as it was decided to develop both the collimator and camera lens barrel systems using in-house facilities. 
\par 
MFOSC-P imaging performance has also been evaluated for its thermal analysis. The typical observing conditions at the site ranges between 5 to 30 degree C for most of the year. At a nominal temperature of 20 degrees C, the chromatic focal range (for 4500-8500$\AA$) is determined to be within $\pm$130$\mu$m. For the circle of confusion to be of 1.1 pixels diameter, the depth of focus is determined to be $\pm$105$\mu$m, which is close to the focus range. For temperature variation between 5 to 30 degree C, the focus shifts (for 4500, 6500 and 8500$\AA$) are found to be within $\pm$130 to 135$\mu$m which is also comparable to the depth of focus. For the given temperature (5 to 30 degree C) and pressure profile at the site, the corresponding RMS diameters are found to be within 16$\mu$m, i.e. $\sim$1.2 pixels, for most of the field points (in imaging mode) and spectral points (along the direction of dispersion in spectroscopy modes). While this slightly overshoots the preferred seeing limited criteria of 1.1 pixels, this performance was deemed acceptable due to a broadened telescope's PSF as explained above. 
\par 
The above range of manufacturing and assembly tolerances and the thermal conditions were incorporated in the final tolerance analysis of the MFOSC-P optical chain using ZEMAX. The extreme range of temperatures variation from -2 to +35 degree C (and corresponding pressure values at Gurushikhar, Mt. Abu site) was considered for this purpose. The statistical performance of simulated systems using ZEMAX Monte Carlo analysis yields the mean of RMS spot diameters within $\sim$1.1 pixels with a standard deviation of 0.3 pixels (for all the modes of imaging and spectroscopy). The $90\%$ of the simulated systems showed the RMS spot diameter values within 1.5 pixels.  The back focal distance of the camera optics was used as a compensator in the tolerance analyses. Its variation was found in the range of $\pm$1.5mm for various cases. This range was primarily useful to provide arrangements for a one-time mechanical movement to the detector system while assembling the instrument in the laboratory. Once the detector is mounted and aligned with the optics, any further change in focus is expected to be due to a change in temperature only, as discussed above.
\par 

\begin{table}
	\centering
	\caption{Image Quality Performance of MFOSC-P optical chain:  RMS spot diameter ($\mu$m)}
	\begin{tabular}{lcccccc}
		\hline        
		\hline                                                                   
		Imaging  &       &       &      &       &     \\ 
		&       &       &      &       &     \\                            
		\hline
		Radial Field Height   & 0  &  1  & 2  &  3  & 4   \\
		(arc-minutes)           &    &     &    &     &     \\
		\hline                            
		Wavelength ($\AA$)     &     &     &    &     &     \\
		\hline
		&       &       &      &       &     \\                  
		4500   & 11.92 & 11.79 & 11.51 & 11.73 & 13.73 \\
		5500   &  9.98 & 10.34 & 11.68 & 13.64 & 15.86 \\
		6500   &  6.50 &  7.00 &  8.79 & 11.41 & 14.42 \\
		7500   &  2.94 &  2.52 &  2.37 &  4.32 &  7.61 \\
		8500   & 12.36 & 11.73 &  9.79 &  6.98 &  4.71 \\
		\hline
		&       &       &      &       &     \\
		Spectroscopy  &       &       &      &       &     \\
		&       &       &      &       &     \\
		\hline
		Grating Mode    & R500  &    & R1000  &    & R2000   \\
		\hline                            
		Wavelength ($\AA$)     &     &     &    &     &     \\
		\hline
		
		4500   &   10.80    &  &   -           &  &   -           \\
		5000   &    7.20    &  &   7.49    &  &   -             \\
		5500   &   10.56    &  &   9.90    &  &   -             \\
		6000   &    9.42    &  &   9.17    &  &   11.52  \\
		6500   &    6.29    &  &   8.06    &  &    5.98         \\
		7000   &    2.62    &  &   8.52    &  &    5.47         \\
		7500   &    2.46    &  &   -           &  &   -             \\
		8000   &    6.16    &  &   -           &  &   -             \\
		8500   &    9.90    &  &   -           &  &   -             \\          
		\hline
		\hline
	\end{tabular}
	\label{table-IQ}
	
	\begin{list}{}{}
		\item Note : Pixel Size is 13 $\mu$m a side.
	\end{list}
\end{table}


\par
\subsection{Opto-Mechanical Design} 
\label{subsec-OptoMechDesign}
\par

A modular approach was adopted while designing the opto-mechanical sub-systems of the instrument. In this scheme, various sub-systems were designed in a way so that they could be assembled and characterized first on the laboratory test bench set-up. These sub-systems were later assembled within the instrument's chassis plate. The collimator and camera optics were to align in their respective barrels as per the assembly-integration-test (AIT) procedure. The lenses were first mounted into their respective lens mounts and were checked for their expected quality on a pre-designed imaging test bench set-up (see section~\ref{subsec-LabCharact}). In this case, simple on-axis optical imaging systems were designed around each of the MFOSC-P lens/lens assemblies using off-the-shelf lenses. The expected imaging performances were derived from the ZEMAX simulations. The image quality for each of the lens assemblies was verified before integrating them into respective barrels. The lens assemblies were aligned and assembled into the barrels using a laser retro-reflection technique. These collimator and camera optics barrels were then assembled into a rigid cage-rod mechanical structure (see Section~\ref{subsec-LabCharact}) to ensure the required opto-mechanical tolerances as per designed AIT procedure.

\begin{figure*}
	\centering
	\includegraphics[width=0.99\textwidth]{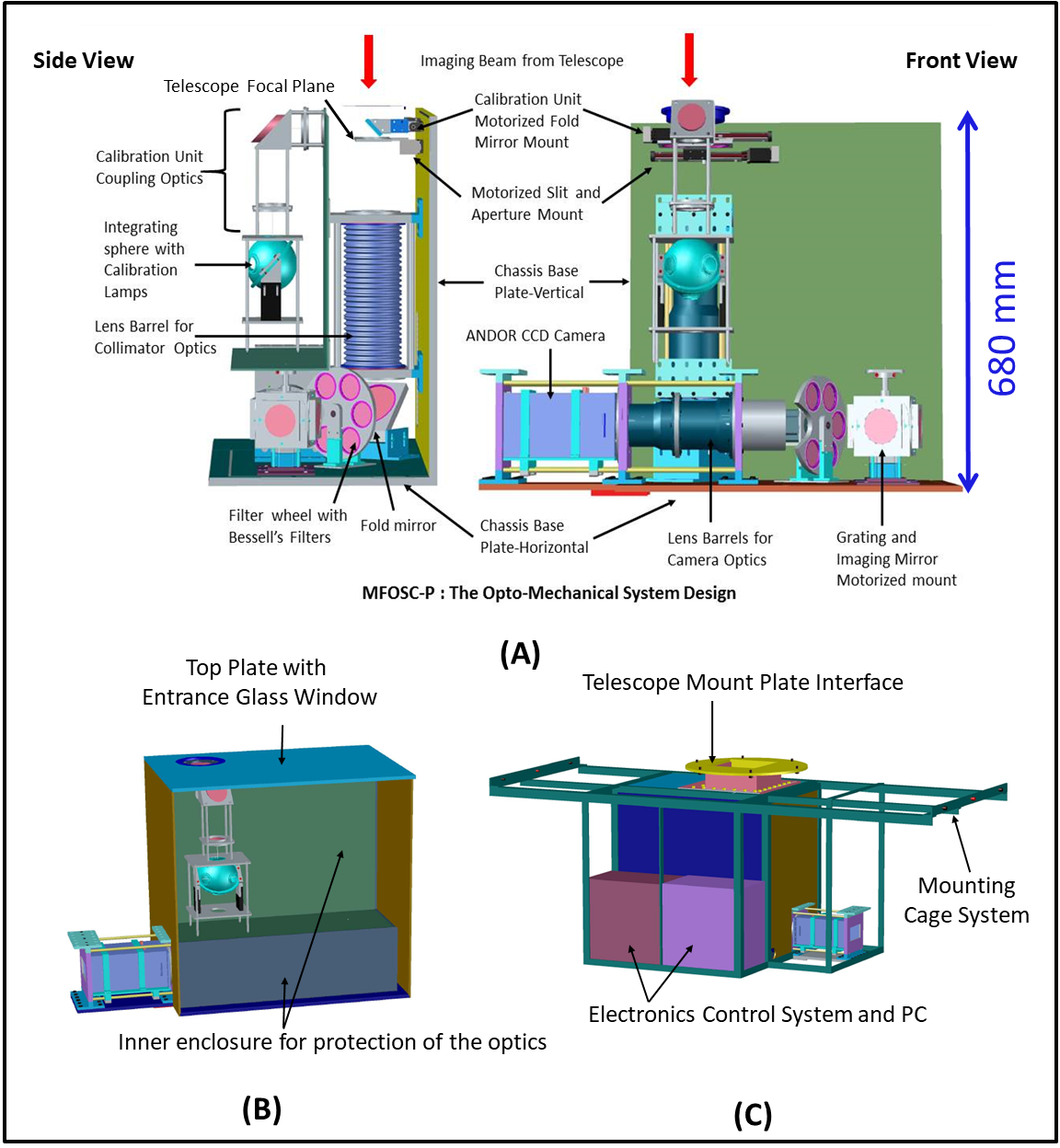}
	\vspace{0.2cm}
	\caption{Opto-mechanical system design of MFOSC-P instrument. Top panel (A) shows the inner design and components of MFOSC-P. Panel (B) shows the outline of the instrument, the inner enclosure for protection of the optics and the outer enclosure. Panel (C) is the final configuration of MFOSC-P with its support structure.}
	\label{fig-MechDesign}
\end{figure*}

\begin{figure}
	\centering
	\includegraphics[width=0.85\textwidth]{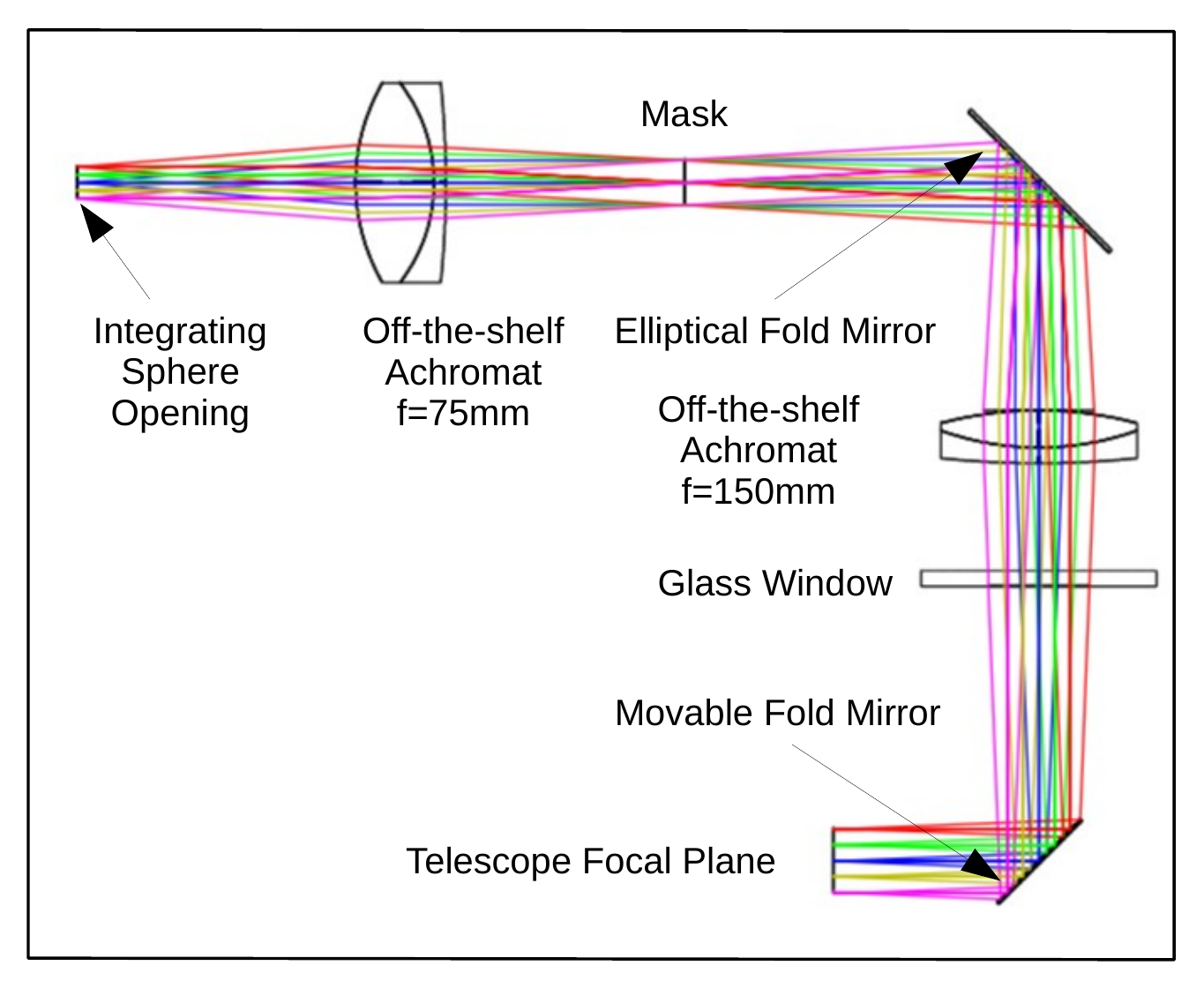}
	\vspace{0.2cm}
	\caption{Optical layout of MFOSC-P calibration unit. The calibration unit is designed with off-the-shelf optical components to simulate telescope's f/13 beam. Telescope pupil is simulated by using a mask. See section~\ref{subsec-OptoMechDesign} for description.}
	\label{fig-CalibUnit}
\end{figure}

\par 
The instrument enclosure is designed in the shape of the box consisting of a 10.0 mm thick L-shape chassis plate. All the opto-mechanical sub-assemblies are mounted on this chassis plate. Figure~\ref{fig-MechDesign} shows the schematics of the opto-mechanical system design of MFOSC-P.  The telescope beam enters into the instrument through an optical protective glass window. The focal plane of the telescope is formed inside the instrument. The slits are positioned on the focal plane using a stepper motor based linear translational stage so that they can be driven in and out of the optics axis. Due to the mechanical design constraints, the field stop is mounted just below this translational stage at the telescope focal plane to limit the FOV and stray light. The collimator optical system then picks the beam from the focal plane and provides a parallel beam space for the filters and gratings requirements. The linear translational stage and collimator optics barrel are mounted on the vertical plate of the L-shape chassis. A 45-degree fold mirror bends the parallel beam in a horizontal plane. The beam passes through a stepper motor-driven filter wheel with Bessell’s B-V-R-I-H$\alpha$ filters (50mm diameter) and an empty slot for spectroscopy purposes. 
\par 
The beam forms the pupil image after the filter on to the grating/imaging mirror surface.  Three different plane reflection gratings and a fold mirror are mounted on the four faces (90 degrees apart) of a rotation mechanism driven by a stepper motor. Desired grating (for spectroscopy) or mirror (for imaging) can be brought into the path of the beam by rotating the mechanism. The dispersed or reflected beam is directed into camera optics and forms the image on the ANDOR CCD detector system. The filter wheel mechanism, gratings motion mechanism, camera barrel, and the detector system are mounted on the chassis's horizontal plate. 
\par
\begin{figure}
	\centering
	\includegraphics[width=0.85\textwidth]{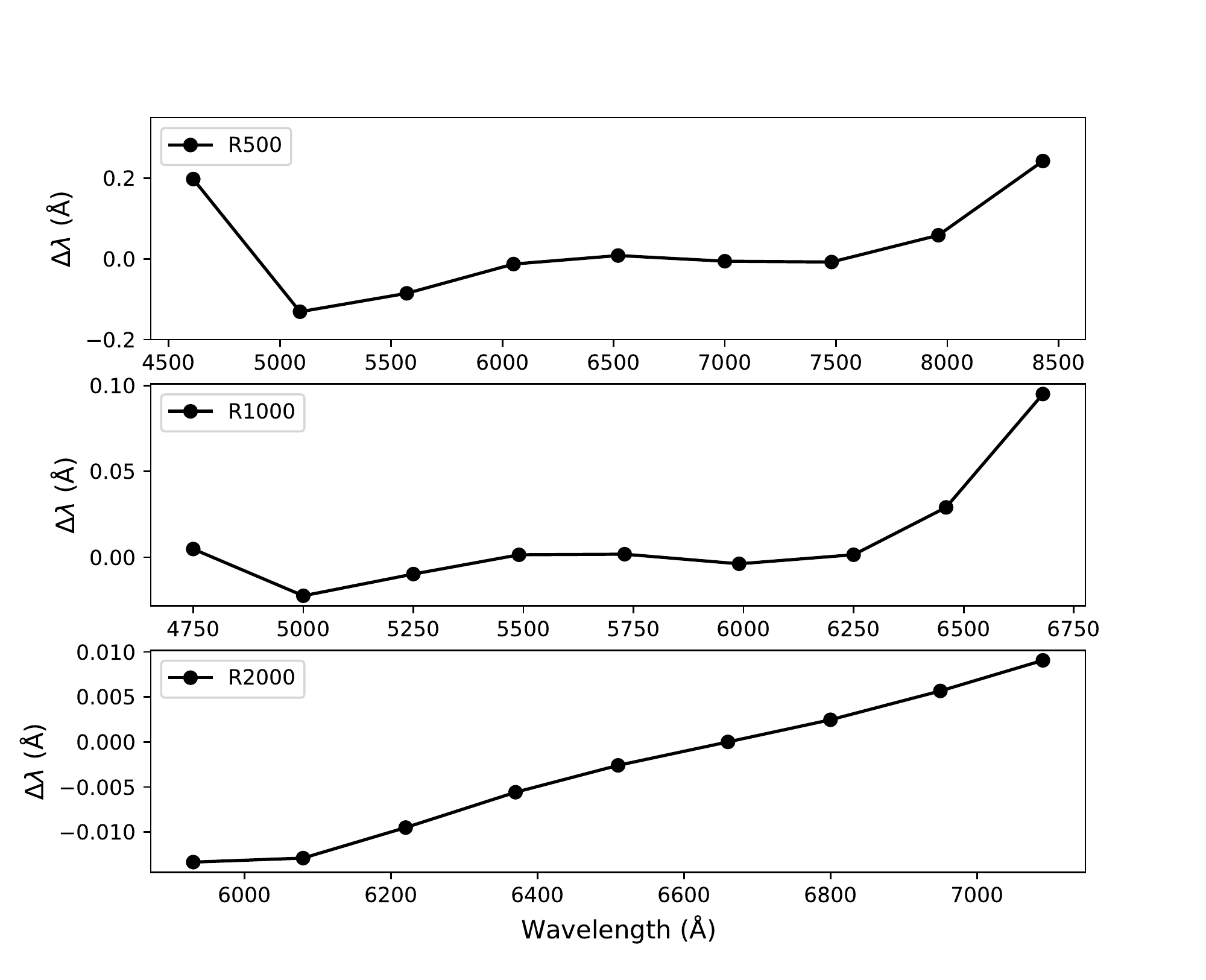}
	\vspace{0.2cm}
	\caption{The expected errors in wavelength calibration, due to the use of calibration unit optics, are shown for all the three grating modes.}
	\label{fig-WaveCalibeError}
\end{figure}

 
Options for wavelength calibration of the spectra have been provided within the instrument. A calibration unit has been designed and integrated with MFOSC-P optics. It consists of an integrating sphere, two spectral lamps (Neon and Xenon), a halogen lamp, and a re-imaging optical system. The pencil-style spectral (Neon and Xenon) lamps from M/S Newport-Oriel are used for this purpose \footnote{https://www.newport.com/f/pencil-style-calibration-lamps, Accessed: 2020-03-18}. The optical layout of the calibration unit is shown in Figure~\ref{fig-CalibUnit}. The lamps are coupled to the in-ports of the integrating sphere. The out-port of the integrating sphere is re-imaged on the focal plane of the telescope with a magnification of $\times$2.0. The re-imaging optics is designed using two off-the-shelf achromat lenses (Thorlabs Part No.: AC508-075-A, 75mm focal length, and Edmund Optics Part No.: 49-391, 150mm focal length) and two mirrors to guide the calibration beam into the instrument. The first fold mirror injects the calibration optics beam into the instrument through a glass window. The second fold mirror is mounted on a stepper motor driven linear translation stage on the vertical chassis plate, above the linear stage for slit movement (Figure~\ref{fig-MechDesign}). This mirror blocks the telescope beam and feeds the calibration beam into the main optics through the slit during the spectral calibration. The second fold mirror is retracted during the science run of the instrument. The telescope pupil is simulated by using a mask (12.0mm clear aperture with a central obscuration of 2.7mm diameter) at the pupil location. Thus, a telecentric $f/13$ beam is generated at the telescope focal plane. The calibration unit optics was simulated with MFOSC-P optical chain in spectroscopy mode to determine the centroid positions for the individual wavelength spots. The centroid positions determined with MFOSC-P $+$ calibration unit optics design are in good agreement with those obtained from MFOSC-P $+$ telescope optics design (within 2-3$\mu$m, i.e., $\sim1/13$ of spectral resolution in higher resolution mode of MFOSC-P). Figure~\ref{fig-WaveCalibeError} shows the expected wavelength errors for all three modes of gratings. The centroids positions of the emission lines of calibration lamps, obtained with the ZEMAX model of calibration unit $+$ MFOSC-P optical chain, are fitted with a third-order polynomial to obtain wavelength solutions. These wavelength solutions were then compared with the centroid positions of individual wavelengths from the ZEMAX model of telescope $+$ MFOSC-P optical chain to derived the expected wavelength errors.

\par 
MFOSC-P is also equipped with an off-axis guider system. Before the beam enters into MFOSC-P, a 45-degree pick-up mirror selects an off-axis field and directly image it on to Lodestar X2 auto-guider camera\footnote{https://www.sxccd.com/lodestar-x2-autoguider; Accessed: 2020-03-19}. The 45-degree pick-up mirror and the Lodestar camera are mounted on a stepper motor based motion mechanism. This mechanism allows movement of the auto-guider system around 90 degrees along the periphery to scan the off-axis field for a suitable guide star.
\par 
The instrument's dimensions are 680mm $\times$ 608mm $\times$ 429mm, and it weighs 68 kg. The opto-mechanical structure is made of aluminum (6061T) and an inner enclosure is added for dust contamination. MFOSC-P is mounted on the telescope along with its control system and control computer using a mild steel support structure (Figure~\ref{fig-MechDesign}). The entire assembly weighs around 121 kg. The finite element analysis of the MFOSC-P mechanical system (including its external support cage system) was also done to ensure its structural integrity and desired tolerances. Iterative analysis and structural modifications were carried out to determine the most optimized configuration of the instrument's structure. The final design configuration was analyzed for self-weight load and acceleration moments. Frequency analysis of instrument (first mode at 24.1 Hz and second mode at 36 Hz) showed that it is safely decoupled with the telescope's structure. Other structural analyses, like self-weight defections in static and dynamic conditions, stress analysis, etc. were performed to ensure compliance with the specified design requirements. The details of structural analysis are described in \cite{Srivastava2018}.

\begin{figure}
	\centering
	\includegraphics[width=0.85\textwidth]{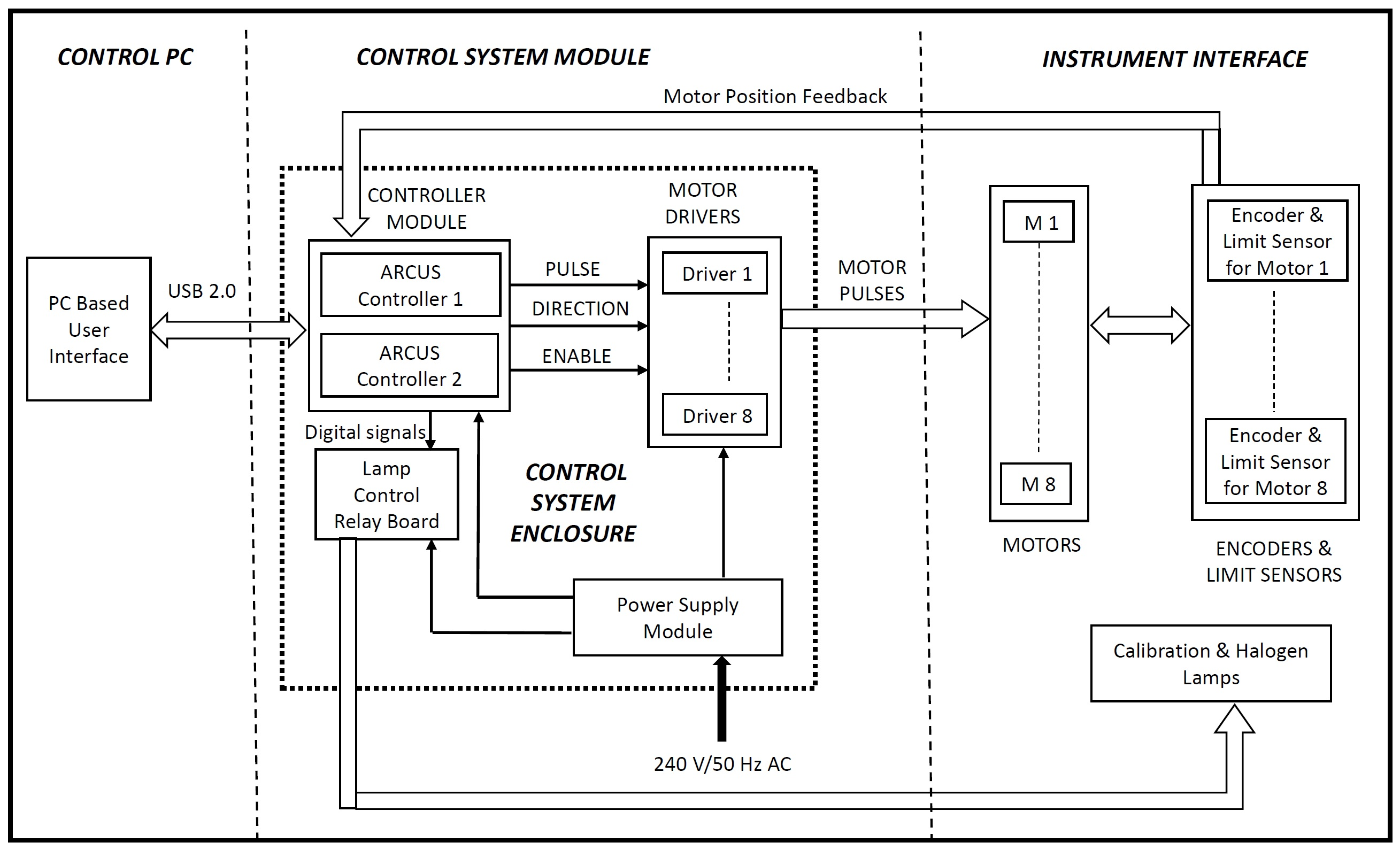}
	\vspace{0.2cm}
	\caption{Block diagram of the hardware architecture of MFOSC-P control system.}
	\label{fig-ControlSys-Hardware}
\end{figure}

\begin{figure}
	\centering
	\includegraphics[width=0.85\textwidth]{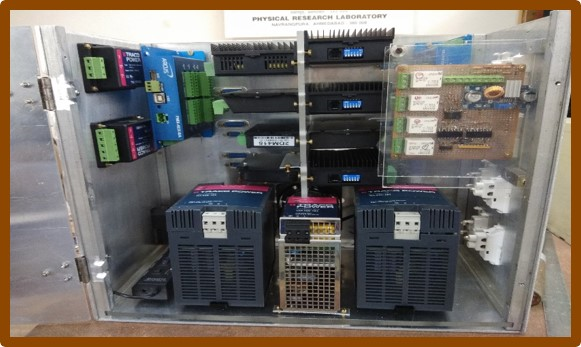}
	\vspace{0.2cm}
	\caption{The hardware set-up of MFOSC-P control system.}
	\label{fig-ControlSystem}
\end{figure}

\begin{figure}
	\centering
	\includegraphics[width=0.90\textwidth]{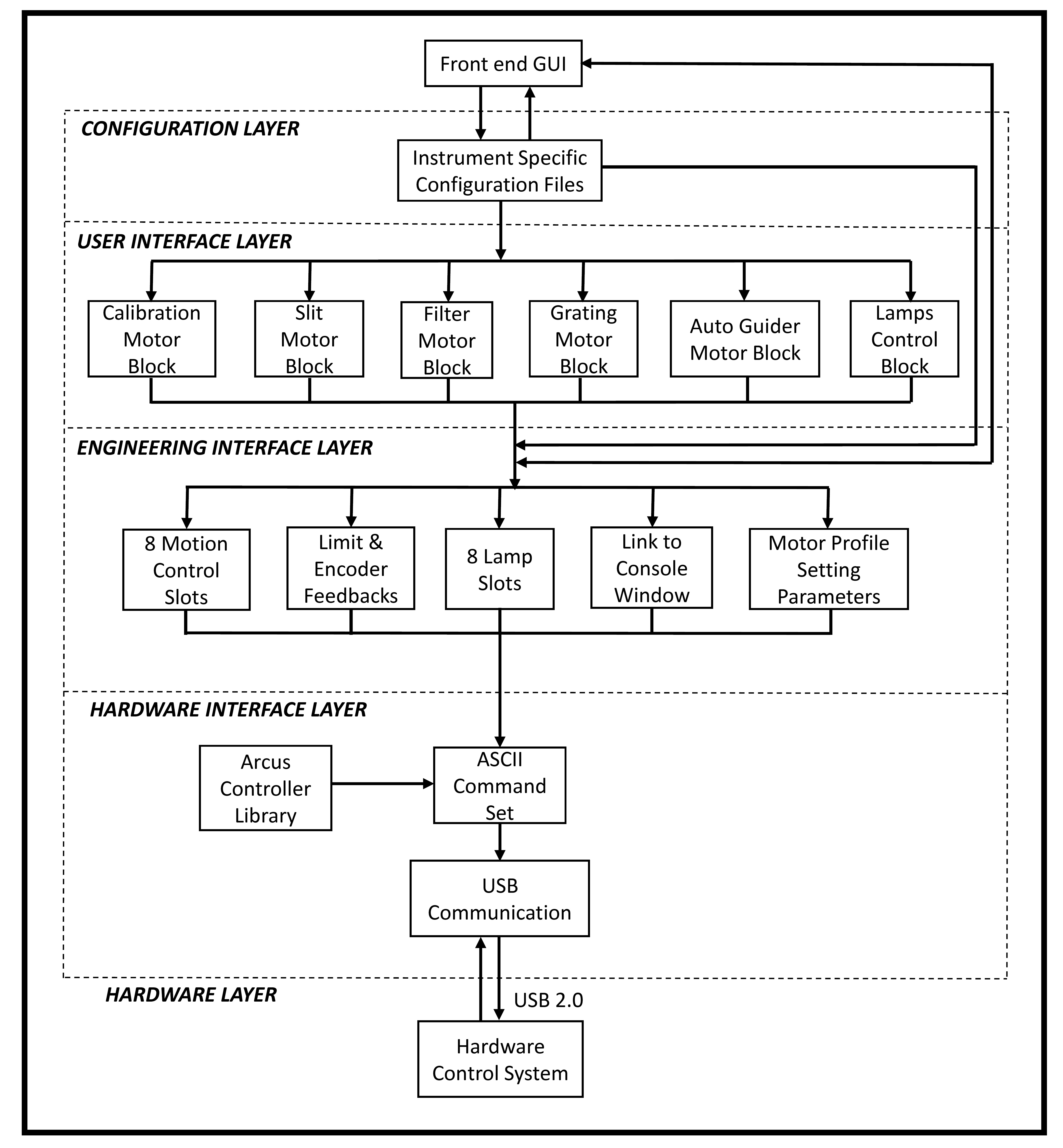}
	\vspace{0.2cm}
	\caption{Block diagram of MFOSC-P control software architecture.}
	\label{fig-ControlSys-Software}
\end{figure}

\begin{figure*}	
	\centering
	\includegraphics[width=0.99\textwidth]{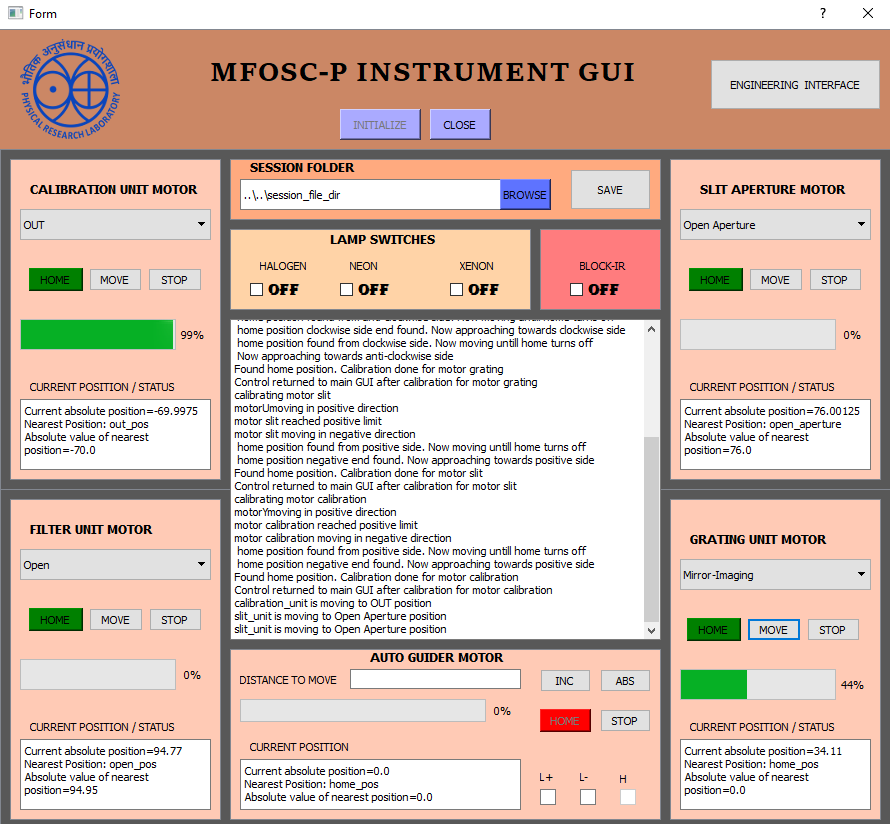}
	\vspace{0.2cm}
	\caption{The graphical user interface (GUI) for MFOSC-P. The GUI has been developed using Python-PyQT framework and acts as the font end of MFOSC-P control software. See section~\ref{sec-ControlSys} for details.}
	\label{fig-GUI}
\end{figure*}

\par 
\section{Automation and Control System}
\label{sec-ControlSys}
\par 

The automation and controls of MFOSC-P are provided in two parts: (1.) The detector system controls, and (2.) the instrument operation controls. The detector system has built-in read-out electronics. It is accessed through the control software - SOLIS - provided by M/S/ Andor Technology Ltd \footnote{https://andor.oxinst.com/products/solis-software/; Accessed: 2020-01-23}. This control software allows a variety of options for detector settings, exposure times, and shutter controls. Different modes of data acquisition can also be selected as per user's requirements, e.g., sub-region selection, different read-out speeds, image spooling and kinetic acquisition, etc. SOLIS also has in-built features for detector cooling and protection. A variety of image processing routines are also available for quick analyses of the data. Other instrument operation controls of MFOSC-P are being provided by an in-house developed instrument control system and graphical user interface (GUI). MFOSC-P has five motion systems, viz. filter wheel, grating motion system, slit $\&$ aperture positioning system, calibration mirror positioning system, and the off-axis auto-guider pick-up mirror. All the motion systems are driven by stepper motors or stepper motor based linear translational stages. Except for the auto-guider system, all other motors are equipped with quadrature encoders. Limits and home sensors are also used in all the motion systems. These motion systems and the lamps in the calibration unit are being operated through the instrument control system.  
\par 
Though off-the-shelf solutions for motion controls are commercially available (e.g., Newport motion controllers\footnote{https://www.newport.com/c/motion-controllers. Accessed: 2019-09-05}), they are generally restricted to two or three axes only. Considering our long-term instrumentation requirements, we decided to develop a general-purpose motion controller using commercially available modules. Such a control system would provide the flexibility of scaling up the number of axes as per the instrument's requirements and would be useful for multiple applications. Figure~\ref{fig-ControlSys-Hardware} shows a schematic block diagram of the control system hardware set-up. The hardware of the control system (Figure~\ref{fig-ControlSystem}) is built around two ARCUS make PMX-4EX-SA series of motion controllers. PMX-4EX-SA is an advanced 4-axes stepper stand-alone programmable motion controller from ARCUS Technology, Inc.\footnote{https://www.arcus-technology.com/ Accessed: 2020-01-23}. It operates on a 24V power supply and can be communicated with USB or RS-485 link. The controller can drive up to four stepper motor drivers, including their encoders and limit/home sensor feedback. It also has the provisions for digital input/output control signals for some auxiliary tasks. In the control system, two ARCUS controllers drive eight commercially available stepper motor drivers. These stepper motor drivers can be individually configured in desired micro-stepping modes and for the required current ratings of the stepper motors. From one of the controllers, four of the digital outputs are used to control the operation of calibration lamps through electro-mechanical relay-based control circuitry. Four different power supplies 24V/10A-DC, 5V/4A-DC, 24V/10A-DC, and 24V/7.5A-DC from TRACO power supplies \footnote{https://www.tracopower.com/home/;Accessed 2020-01-23} are used for controller operations, relay circuit operations, limit sensor operations and motor operations, respectively. While their input power is derived from mains 240V/50Hz AC, all these four power supplies are optically isolated with each other in the control system.  
\par
The GUI software has been designed and developed to serve two functionalities : (1.) To use the control system as a general-purpose 8-axes motion control system for multiple purposes (2.) MFOSC-P specific user's interface. GUI and its application software have been developed using Python 2.7 with the PyQt4 framework. Figure~\ref{fig-ControlSys-Software} describes the block level architecture of the application software. The front facade is the main front-end GUI window, which is designed as per the requirements of the MFOSC-P instrument (see Figure~\ref{fig-GUI}). It provides the visualization and status of various motion sub-systems. The progress of any motion system while changing modes can be monitored on the GUI screen. The front layer takes the user's inputs/tasks. It then communicates to a general-purpose engineering interface through an instrument-specific configuration file and sub-system specific user interface layer. The configuration file is stored in an editable ASCII text format and contains the default set-up status of various sub-systems of MFOSC-P. It acts as the interface between the instrument-specific main GUI screen and controller specific engineering interface. Provisions are made in the front-end GUI to access the engineering interface directly to access the hardware component (e.g., individual motors, limit sensors, lamp control, etc.) separately for engineering tests and debug purposes. The engineering interface is password protected and has another set-up window to store each motor's motion profile parameters (e.g., start speed, top speed, acceleration, deceleration, pitch, micro-stepping mode, etc.). The engineering interface window also provides access to a command-line console interface, where one can directly interact with the controllers using the ASCII command sets provided by controller's manufacturers in the form of the library functions. A USB communication layer then forms and sends the command packet to the hardware system using the USB2.0 interface.
\par
The software is designed to be platform-independent and easily scalable. The main front window of the GUI is designed and configured as per MFOSC-P requirements. However, the main GUI screen and configuration file can be appropriately changed to make the software control suitable for any other future instrumentation requirement while keeping the control system hardware intact. \\

\par 
\section{Characterization and Commissioning Performance} 
\label{sec-Charactrization}
\par 

MFOSC-P was assembled and characterized in the laboratory from October 2018 to January 2019. Performance verification of the optics components (e.g., lenses and lens sub-assemblies), motion sub-systems testing, imaging and spectroscopy performance verification, etc. were done in this period. Subsequently, MFOSC-P was commissioned on the telescope in February 2019. Regular characterization and preliminary science observations were made from February 2019 till June 2019. The laboratory and on-sky performance are described in the next sub-sections.

\par
\subsection{Laboratory Characterization} 
\label{subsec-LabCharact}
\par

\par
\subsubsection{Image Quality} 
\label{subsubsec-LabIQ}
\par

\begin{figure*}
	\centering
	\includegraphics[width=0.99\textwidth]{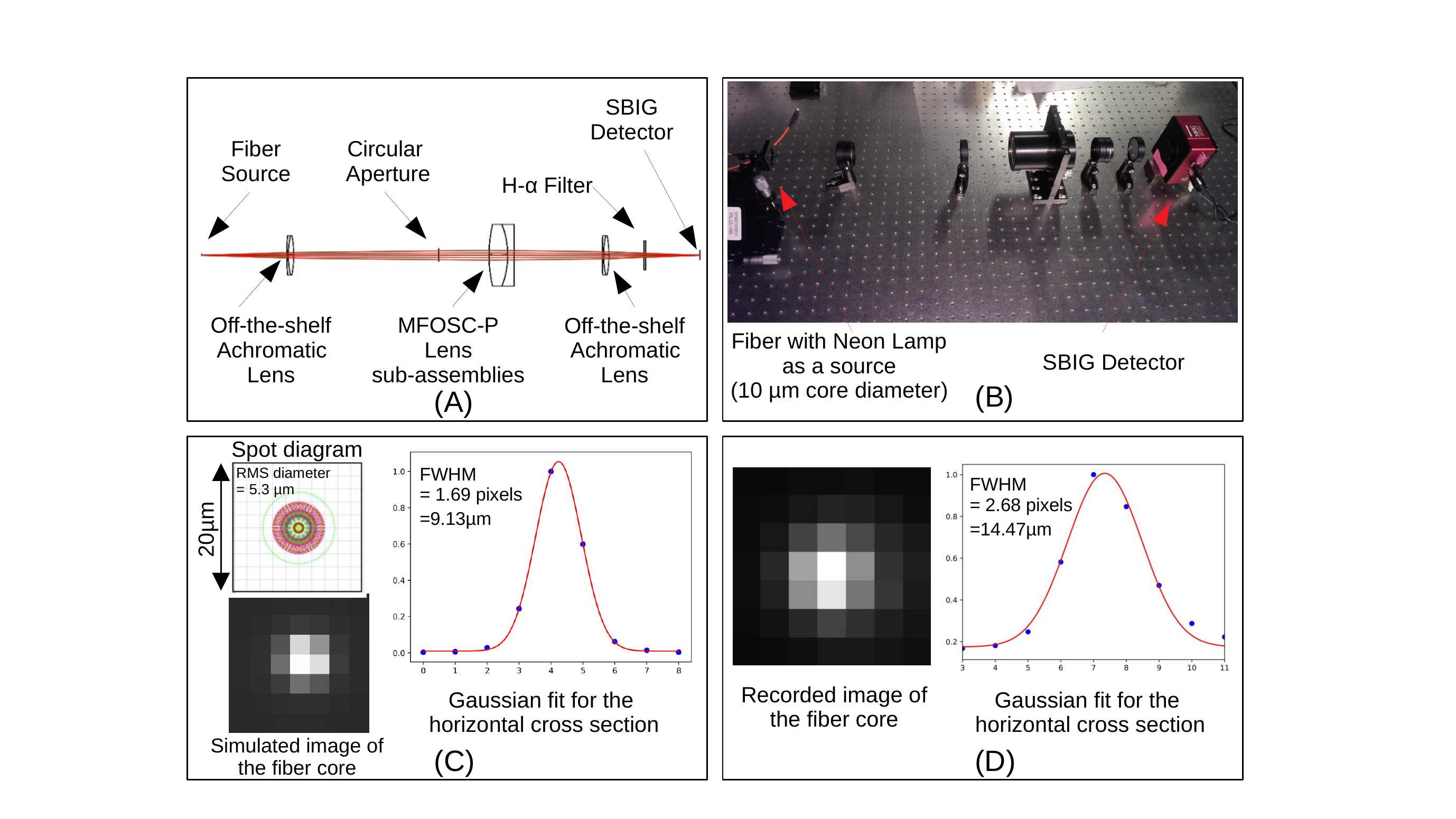}
	\vspace{0.2cm}
	\caption{The figure shows a test set-up designed to verify the basic optical performance of individual lenses and lens assemblies. Panel (A) shows the ZEMAX optical layout of the test set-up. This set-up was designed to provide magnification of $\times$0.8 so that a fiber core diameter of 10$\mu$m would fall within 1.5 pixels of the detector. Panel (B) shows the laboratory set up as per the ZEMAX design. The spot diagram, simulated image and expected profile of the fiber core are shown in panel (C). The recorded image of the fiber core and resultant profile are shown in panel (D). SIBG camera with 5.4$\mu$m pixel size was used. The measured PSF was in good agreement with the predicted ZEMAX performance. Similar set-ups were designed and verified for each of the lenses and lens assemblies of MFOSC-P.} 
	\label{fig-LensCheckSetup}
\end{figure*}

The lenses were manufactured as per design by a commercial optics manufacturer (M/S Gooch and Housego PLC, UK \footnote{https://gandh.com/; Accessed: 2020-03-20}) and later assembled into the collimator and camera barrels in the laboratory. The lenses were assembled into the barrels in a two-step process. First, each of the lenses was mounted to their respective mounts. In some cases, two lenses were designed to be kept very close to each other. These lenses were housed in a single mount, thereby forming a lens system sub-assembly.  These mounts and sub-assemblies were later housed in the barrels. Before the fabrication of the barrels, the primary optics specifications (e.g., focal length) and lens quality of each of the lenses were checked by designing individual test bench set-up for each of the lenses (or in some cases for lens sub-assemblies) separately. Different on-axis imaging systems were designed for the test bench around each of the MFOSC-P lens/lens assemblies. Commercial off-the-shelf achromat doublet lenses were used for this purpose. An optical fiber, of 10$\mu$m core diameter, was illuminated with Neon lamp at one side. The output end of the fiber was imaged using the test set-up on a laboratory quality SBIG camera system (Model No. STF-8300M, 5.4$\mu$m pixel size) with the pre-determined magnification ratio. Narrow-band H-$\alpha$ filter and a small aperture on pupil were also used in the set-up design to reduce chromatic and geometric aberrations. The image quality was then measured and compared with the expected performance determined with ZEMAX software. Figure~\ref{fig-LensCheckSetup} shows the one such set-up to verify the lens specification before its assembly into the barrel. Though such schemes would only provide a good on-axis image quality, they are found to be very useful in verifying the coarse specifications of the lenses, in the absence of a costly wavefront analysis set-ups like Interferometer.


\begin{figure*}
	\centering
	\includegraphics[width=0.98\textwidth]{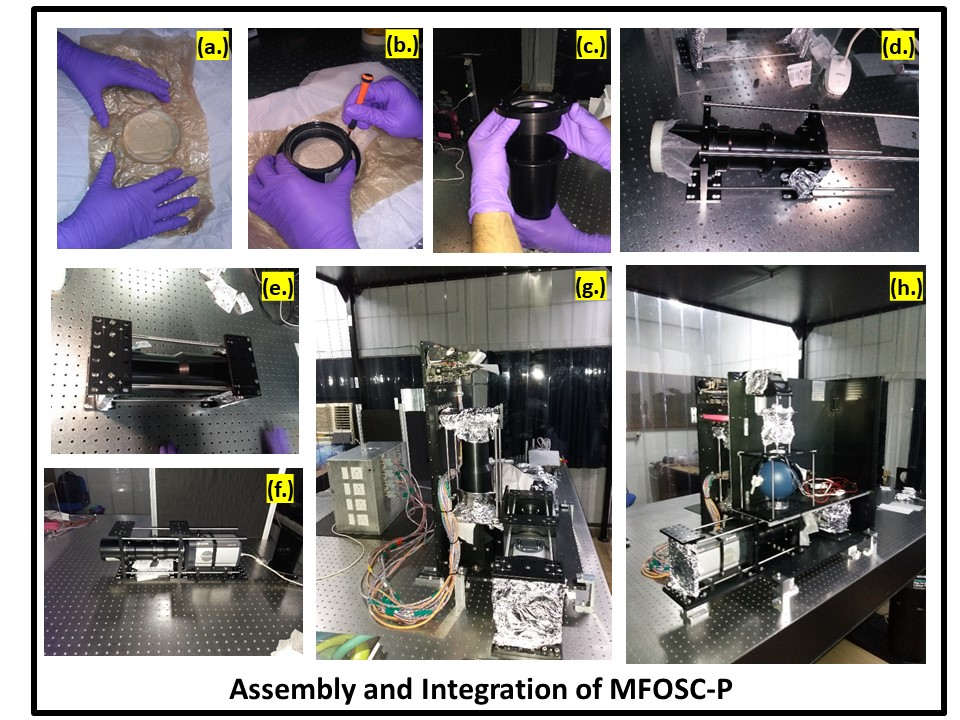}
	\vspace{0.2cm}
	\caption{ The figure shows some aspects of assembly and integration of MFOSC-P. Panels (a), (b) and (c) show the development of lens mounts and barrels. Panel (d) shows the camera optics barrel under during development. The assembled lens barrels including the optics for collimator and camera are shown in panels (e) and (f). The final assembly of the instrument is shown in panels (g) and (h) without outer enclosure cover plates.}
	\label{fig-AIT}
\end{figure*}

\begin{figure}
	\centering
	\includegraphics[width=0.98\textwidth]{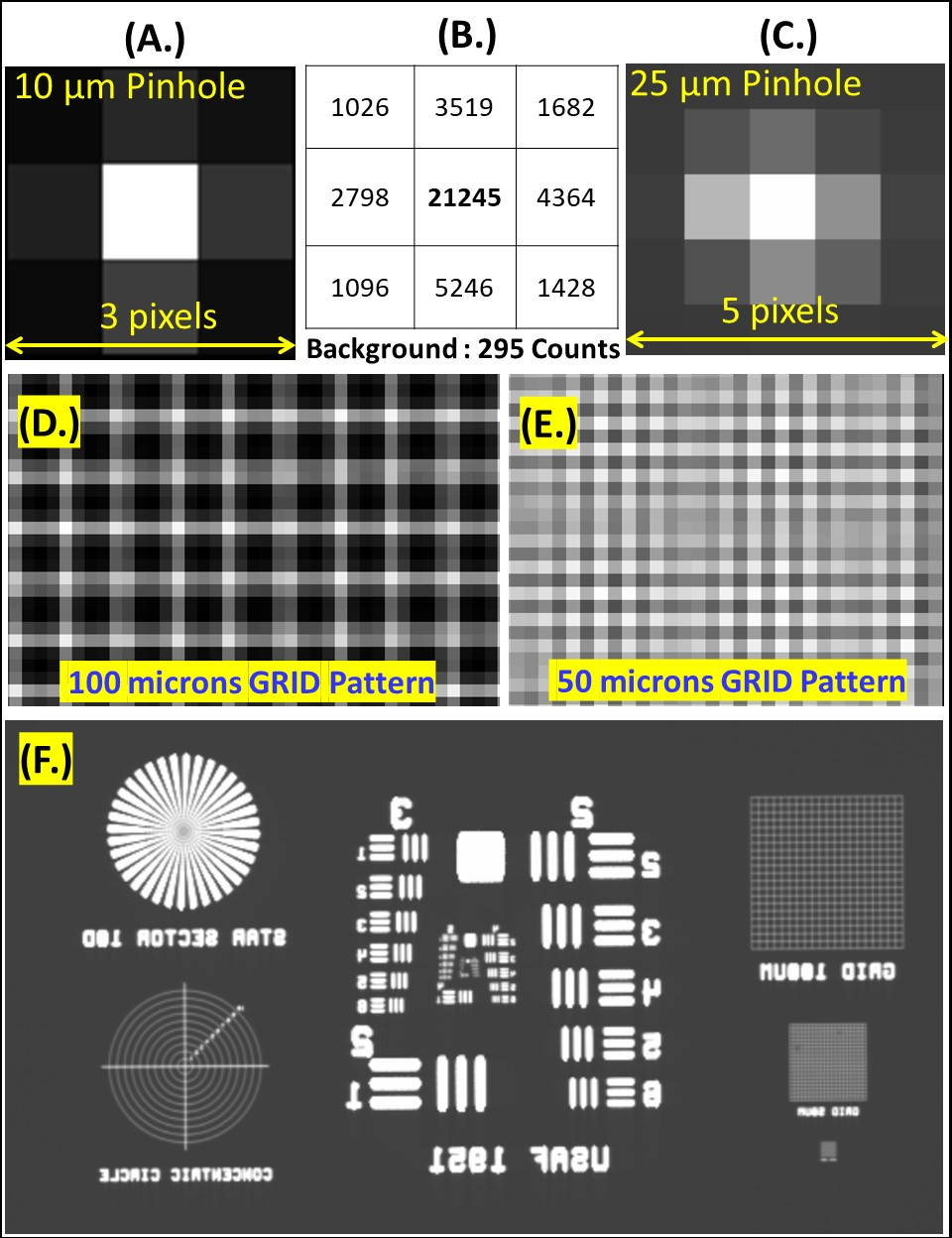}
	\vspace{0.2cm}
	\caption{Image quality characterization of MFOSC-P. Panel (A) displays the footprint of 10$\mu$m pinhole illuminated with Neon lamp on the MFOSC-P CCD detector. Panel (B) shows the counts recorded in central 3$\times$3 pixels for 10$\mu$m pinhole. The mean background counts are 295. The FWHM of the PSF is estimated to be $\sim1.16$ pixels. Panel (F) shows the images of the standard test and resolution target. It has grid structures of 100$\mu$m and 50$\mu$m, which would correspond to pitch of 4.4 and 2.2 CCD pixels, respectively. Panels (D) and (E) show these grid patterns. The grid patterns are well resolved by MFOSC-P optics.} 
	\label{fig-IQ_PSF}
\end{figure}

\par  	
The mounted lenses were then assembled into collimator and camera optics barrels using the laser retro-reflection method. ANDOR camera system was integrated with the camera optics barrel assembly. Figure~\ref{fig-AIT} shows some of the aspects of the MFOSC-P AIT process. These optics sub-systems were evaluated for the image quality assessment on the laboratory optical test bench before integrating into the instrument chassis. The calibration unit optics (discussed in Section~\ref{subsec-OptoMechDesign}) with Halogen, Xenon, and Neon lamps were used to simulate the f/13 beam on the test bench. The image quality of the MFOSC-P optical chain was characterized on the test bench with off-the-shelf pinholes of various diameters (10, 25, 50 $\mu$m, etc.) and resolution test target from M/S Thorlabs Inc\footnote{https://www.thorlabs.com; Accessed: 2020-03-20} in transmission mode. The pin holes were mounted on 2-D translational stages to move it over the face of the CCD detector. As MFOSC-P optics provides magnification of $\times$0.57, 10$\mu$m pinhole mapped onto 0.44 pixels of the ANDOR camera system. Figure~\ref{fig-IQ_PSF} shows some of the images obtained to estimate the image quality of MFOSC-P optics. Panels (A) and (B) show the intensity pattern of 10$\mu$m pinhole illuminated by Neon lamp without any filters. A 2-D Gaussian fits this pattern with FWHMs of 1.36 and 1.40 pixels in the X and Y directions.  A simple quadrature summation method is used to estimate the Point Spread Function (PSF) of MFOSC-P optics. Thus, the geometrical projection of the pinhole is removed from the measured FWHMs. The optics PSF is determined as a Gaussian with FWHM of 1.16 pixels, which corresponds to $\sim$80$\%$ encircled energy within 1.5 pixels. This result was consistent with the results of opto-mechanical tolerance analysis, as discussed in section~\ref{subsec-IQandTol}. The PSF was then determined, for each of the B-V-R-I optical filters, by imaging the pinhole at different coordinates on the face of the detector. The measured PSFs are in good agreement with the predicted performance of the optical design with tolerance analysis. The image quality was also determined by using off-the-shelf resolution and distortion test targets from M/S Thorlabs (Part No. R1L1S1N). The 100$\mu$m and 50$\mu$m grid patterns, corresponding to the pitch of 4.4 and 2.2 pixels of the CCD respectively, are well resolved by the MFOSC-P optics chain. The image of the test target is shown in the panel (F) of Figure~\ref{fig-IQ_PSF}.

\par 	
\subsubsection{Spectroscopy Performance and Resolutions}
\label{subsubsec-LabSpec}
\par

\begin{figure}
	\centering
	\includegraphics[width=0.98\textwidth]{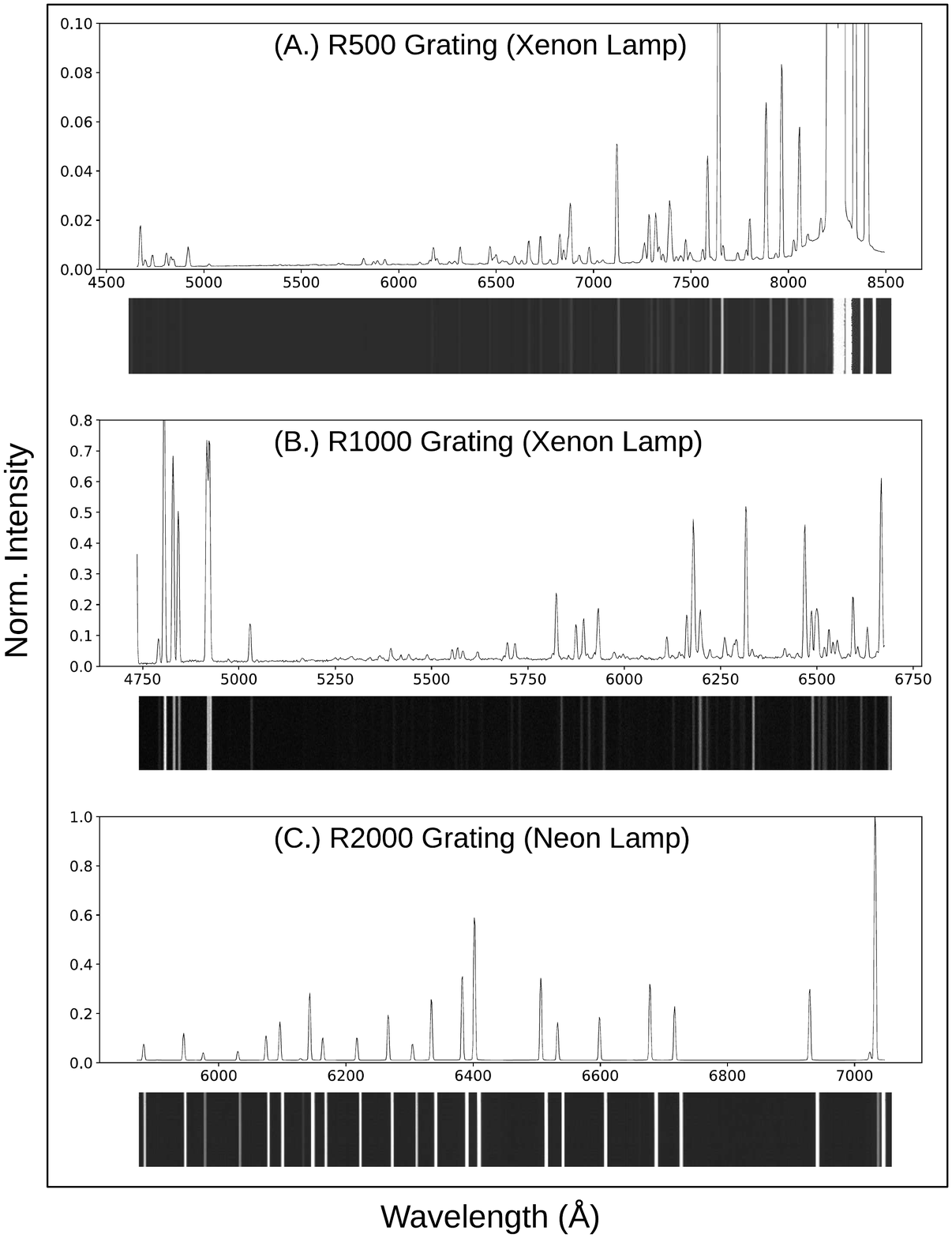}
	\vspace{0.2cm}
	\caption{Spectra of Xenon and Neon lamps used for wavelength calibration in MFOSC-P.}
	\label{fig-LampSpec}
\end{figure}

\begin{figure}
	\centering
	\includegraphics[width=0.98\textwidth]{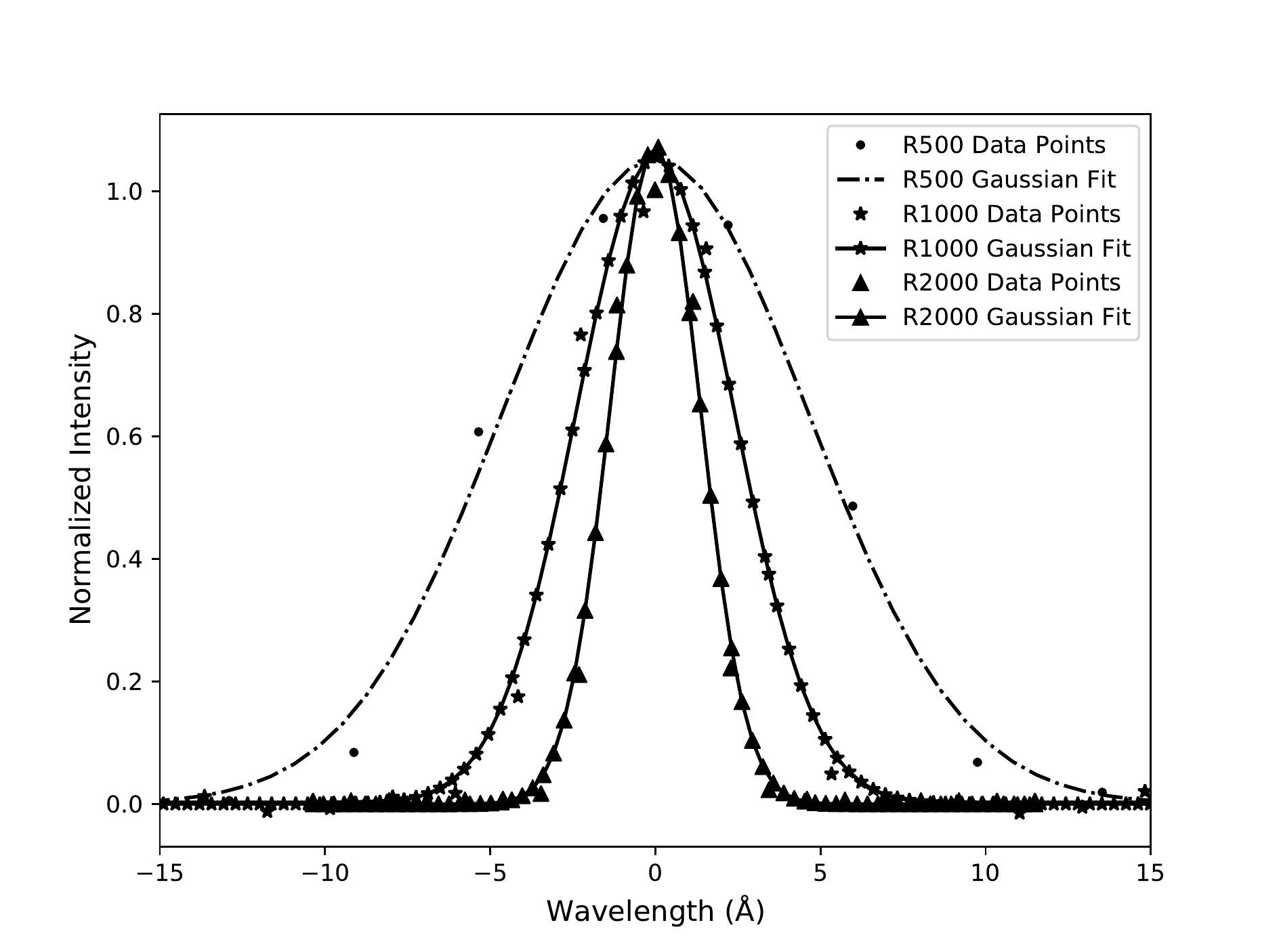}
	\vspace{0.2cm}
	\caption{Line profiles for Xenon and Neon spectral calibration lamps are shown. The central zero wavelength corresponds to 	Xenon lamp's 7119$\AA$ line using R500 grating (Dots), Xenon lamp's 5028$\AA$ line using R1000 grating (stars), and Neon lamp's 6506$\AA$ line using R2000 grating (triangles). Gaussian fits to these lines result in $\sim$3 pixels FWHMs which correspond to 10.9$\AA$, 5.7$\AA$ and 3.2$\AA$ for R500, R1000, and R2000 modes respectively.}
	\label{fig-LineProfile-Joint}
\end{figure}

MFOSC-P optics was also characterized in all the three spectroscopy modes using lamps spectra with the calibration optics set-up. Figure~\ref{fig-LampSpec} shows the images and reduced spectra of Xenon and Neon spectral lamps using the slit of 75$\mu$m width.  In all the cases, emission line profiles are well approximated by the Gaussian function with FWHM of $\sim$3 pixels, which corresponds to 10.9, 5.7, and 3.2 $\AA$ for R500, R1000 and R2000 grating modes, respectively. Figure~\ref{fig-LineProfile-Joint} shows the profiles of these emission lines for R500, R1000, and R2000 grating modes. Wavelength solutions for all the three gratings are determined by using such spectra of the calibration lamps. A third-order polynomial fit is used to determine the wavelength solutions.


\begin{figure*}
	\centering
	\includegraphics[width=0.99\textwidth]{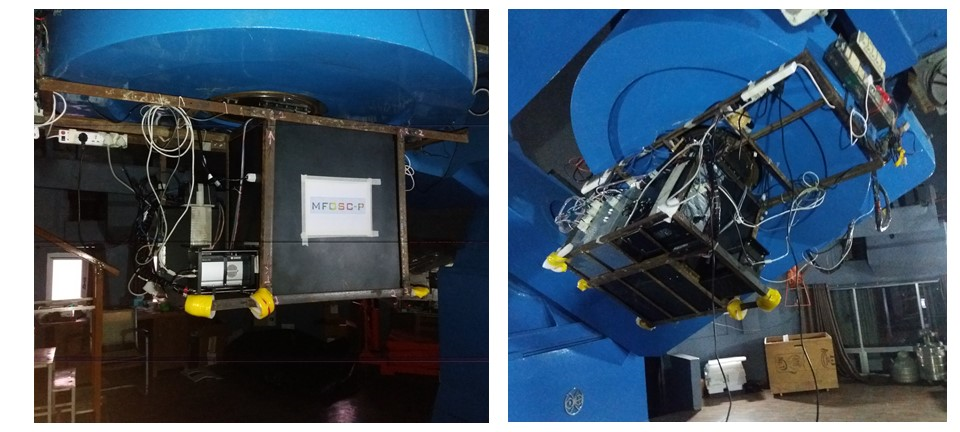}
	\vspace{0.2cm}
	\caption{MFOSC-P instrument mounted on the PRL 1.2m Telescope.}
	\label{fig-OnTel}
\end{figure*}


\par
\subsection{On-Sky Characterization} 
\label{subsec-OnSkyCharact}
\par

Several commissioning and science observations were made with MFOSC-P from February to June 2019; till the onset of the monsoon season during which the observatory is closed. A variety of targets were observed over these five months and later after the monsoon season to ensure and confirm the performance of spectroscopy and imaging modes. Figure~\ref{fig-OnTel} shows the final mounting configuration of the MFOSC-P instrument on the PRL 1.2m Telescope.

\begin{figure*}
	\centering
	\includegraphics[width=0.98\textwidth]{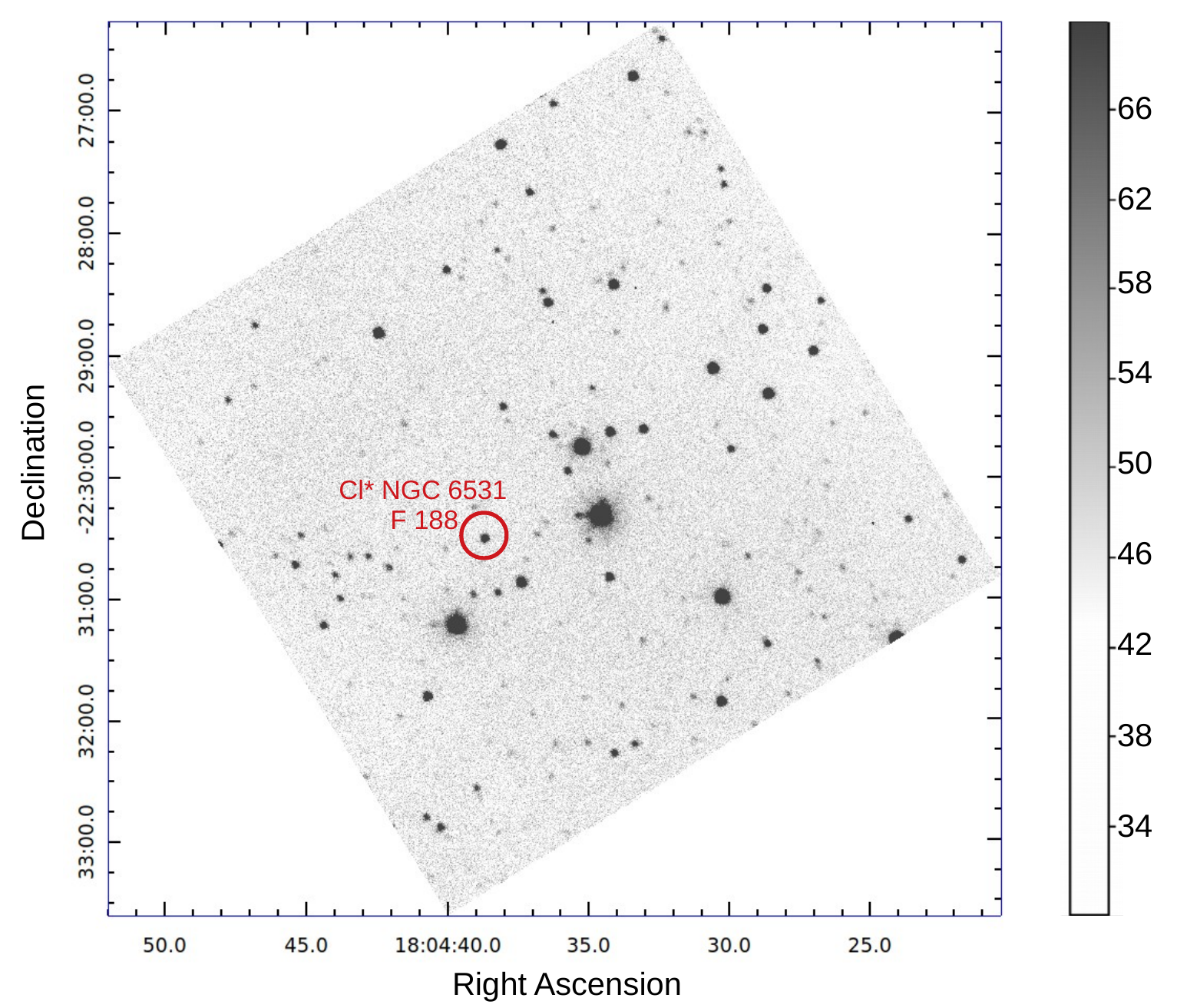}
	\vspace{0.2cm}
	\caption{Images of a region of M21 Open Cluster obtained by MFOSC-P in V filter. A source of V magnitude $\sim$15.74 (encircled) is observed with an error of 0.15 magnitude (SNR$\sim$7.3) in 40 seconds of integration time.}
	\label{fig-Imaging-V-Band}
\end{figure*}

\begin{figure}
	\centering
	\includegraphics[width=0.98\textwidth]{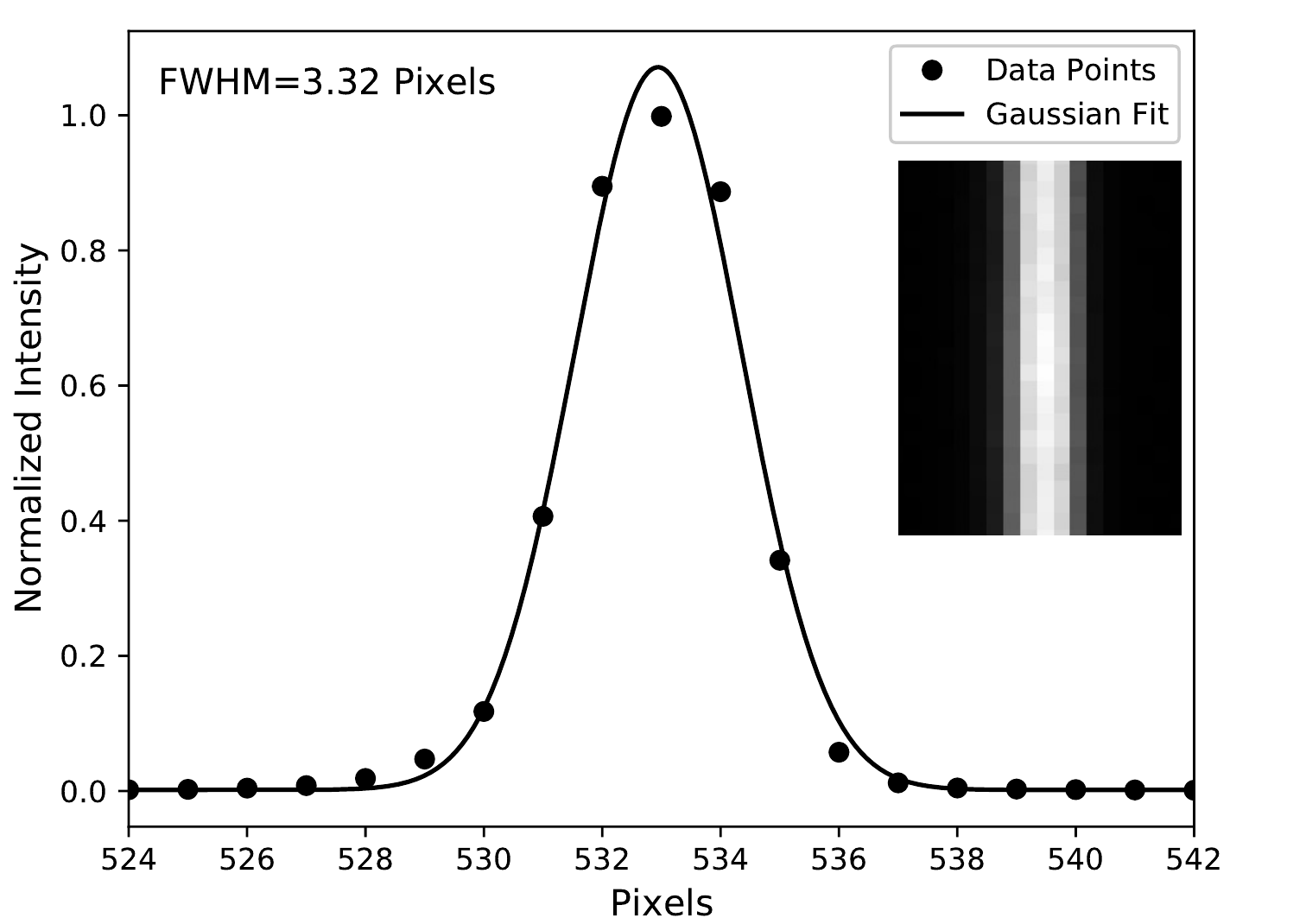}
	\vspace{0.2cm}
	\caption{Slit profile of the 75$\mu$m slit recorded with diffused illumination on the telescope dome floor. The instrument was mounted on the telescope. The profile is well approximated by a Gaussian having FWHM of $\sim$3.3 pixels. A section of the slit image is also shown in the inset.}
	\label{fig-SlitProfile-DiffuseIllum}
\end{figure}


\par
\subsubsection{Photometry} 
\label{subsubsec-SkyPhot}
\par

The imaging PSF of MFOSC-P is heavily dominated by telescope optics along with the seeing profile of the site. On a typical night, the PSF was found to be having FWHM of $\sim2.3$ arc-seconds (measured using a separate CCD camera). Figure~\ref{fig-Imaging-V-Band} shows the image of a region of M21 open cluster observed in the V band filter of MFOSC-P. The magnitude of one of the faint sources (encircled in Figure~\ref{fig-Imaging-V-Band}) is determined to be 15.74 with the photometric error of 0.15 magnitude (corresponding to SNR of 7.3) in 40 seconds of integration time.
\par 
As the telescope PSF was not good enough to check the instrument's PSF on the telescope, we took the image of the 75$\mu$m slit in diffused dome lights (with MFOSC-P mounted on the telescope) without any filter. The slit profile is shown in Figure~\ref{fig-SlitProfile-DiffuseIllum}. A Gaussian fit results in FWHM of 3.3 pixels, which ensure the intrinsic image quality of MFOSC-P on the telescope. 


\begin{figure*}
	\centering
	\includegraphics[width=0.85\textwidth]{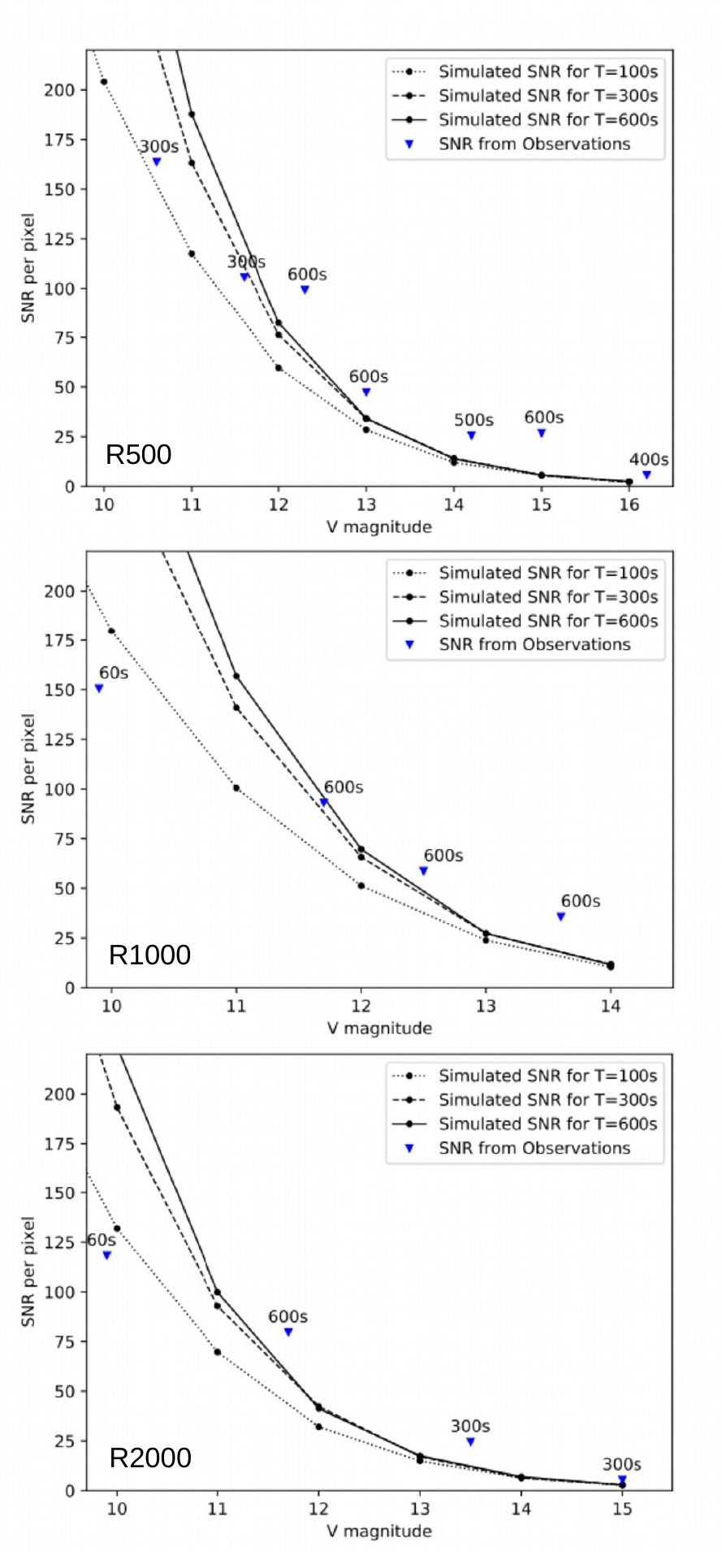}
	\vspace{0.2cm}
	\caption{The plots show the expected SNRs from MFOSC-P for various V band magnitudes, for all the three spectroscopy modes. The expected SNR (solid circles) are obtained from a theoretical simulation model for 100 seconds (dotted line), 300 seconds (dashed line), and 600 seconds (solid line) of the simulated observations. The achieved SNR from MFOSC-P observations for varying integration time, and various magnitudes are also shown in the solid triangles. The corresponding integration times for each of the observations are also mentioned, along with the data points. The achieved SNRs are in good agreement with the predicted SNR values. See section~\ref{subsubsec-SkySpec} for details.}
	\label{fig-SNR}
\end{figure*}


\par
\subsubsection{Spectroscopy Performance} 
\label{subsubsec-SkySpec}
\par

A variety of objects, e.g., symbiotic systems, novae, M dwarfs, etc. were observed with MFOSC-P over the last couple of years. There are described later in this section. The instrument's sensitivity was verified by comparing SNR obtained with the MFOSC-P spectra of these sources with a theoretical simulation model. The theoretical model assumes a seeing of 2.0 arc-seconds FWHM, slit width of 1.0 arc-second, and the optics PSF as a Gaussian of 1-pixel FHWM. A flat V band spectrum was then simulated, incorporating the total instrument efficiency (see Figure~\ref{fig-Throughput}). The spectrum was convolved with the slit profile function and then digitized using the gain of the CCD. The spectrum was finally recorded against a laboratory recorded bias frame, including the read-out noise, dark noise, etc. This simulated spectrum was then extracted using the same aperture and algorithm, as in the case of MFOSC-P observed spectra. The SNRs were determined for each pixel after binning the data along the cross-dispersion direction within the aperture. Figure~\ref{fig-SNR} shows the SNR per pixel achieved with all the three modes of MFOSC-P against their predicted values for sources of various magnitudes and exposure times. The achieved SNRs are in good agreement with the predicted values. In cases where the achieved SNR differs from the expected values, we expect varying seeing conditions and the grating efficiency to be the main reasons for the difference. The efficiency curves used in these simulations were obtained from the manufacturer's datasheet. The curves were generated for the littrow configuration, which is different from the configurations used within MFOSC-P. The stability of the instrument was verified by undertaking a long-term science observing program of M Dwarfs after the instrument was commissioned. Various subtypes of M Dwarf covering a range of magnitudes were observed between February-June 2019. The spectra obtained were then compared with their respective template spectra and later analyzed for their expected outcomes. This program is discussed later in this section.   
\par
\begin{figure*}
	\centering
	\includegraphics[width=0.99\textwidth]{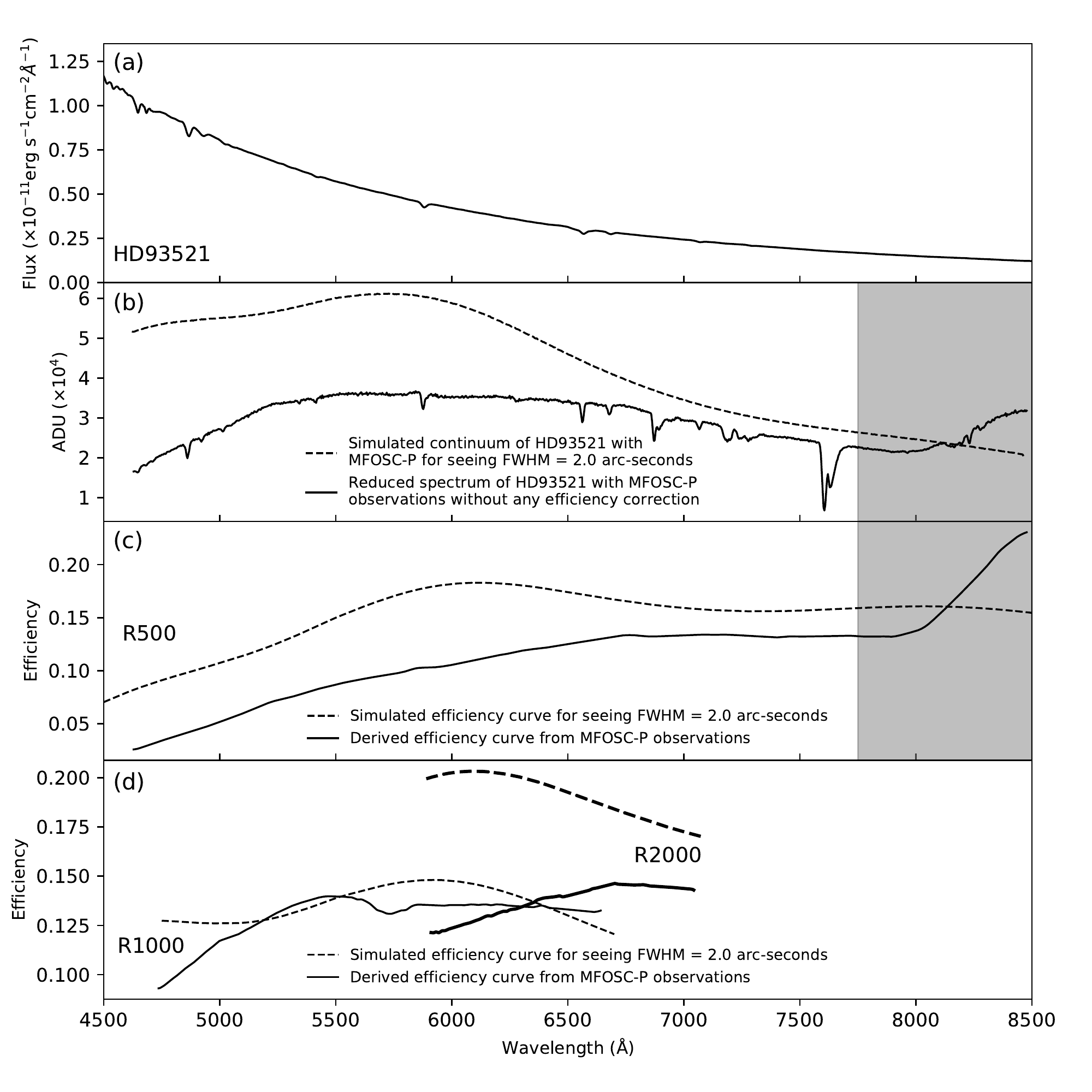}
	\vspace{0.2cm}
	\caption{Plots showing the on-sky efficiencies of the MFOSC-P as derived from the observations. Panel (a) shows the flux calibrated spectrum of a spectro-photometric standard star HD93521. Panel (b) shows the observed spectrum of HD93521 with MFOSC-P R500 mode (solid line). The data is in units of analog to digital units (ADU) as obtained from the CCD detector without any efficiency correction. Dashed curve shows the simulated spectrum (only continuum) of the star with MFOSC-P considering the seeing of 2 arc-seconds.  Panel (c) and (d) show the expected efficiency using simulated data (dashed line) and derived efficiency using observations (solid line) of the instrument for given seeing conditions for R500 and R1000/R2000 modes respectively. The shaded regions in panel (b) and (c) show enhancement due to second order contamination. See section~\ref{subsubsec-SkySpec} for discussion.}
	\label{fig-OnSkyThroughput}
\end{figure*}

 
The on-sky throughput of MFOSC-P, including telescope optics and seeing, is estimated by observing the objects from the list of spectro-photometric standard stars\footnote{https://www.eso.org/sci/observing/tools/standards/spectra.html, Accessed: 2021-02-21}. In addition to the instrument's intrinsic efficiency, several other factors govern the final throughput. As discussed above, the PSF at the telescope's focal plane is dominated by the telescope optics and the seeing profile, which causes the slit-loss. The reflectance of the telescope optics (primary and secondary mirror) also degrades over time since their aluminization. The grating efficiency curves provided by the manufacturer is for the littrow configuration, which differs from the way gratings are being used in MFOSC-P.  We have incorporated the combined effects of all these factors into the final "seeing" PSF on to the MFOSC-P slit. Similar to what we have done for comparing the SNR (in the above paragraph), we have simulated the spectra of a standard star HD93521 ($V \sim 7$ magnitude), with the efficiency curves discussed in section~\ref{subsec-IQandTol} and for the seeing of 2.0 arc-seconds. This simulated spectrum was then compared with the observed MFOSC-P spectrum to deduce the instrument efficiencies. Figure~\ref{fig-OnSkyThroughput} shows the flux calibrated spectrum of HD93521 in panel (a). The observed (solid curve) and simulated spectra (dashed curve)) for R500 mode (for an integration time of 30 seconds) are shown in panel (b) in analogue-to-digital units (ADUs), without any corrections. Only the continuum of HD93521 is simulated. The derived and simulated efficiencies for R500 mode (solid and dashed curve, respectively) are shown in panel (c) for comparison. The above process is also repeated for R1000 and R2000 modes (for an integration time of 45 and 80 seconds, respectively) and their efficiencies are compared in panel (d). It is seen that the efficiency curve is well matched for the R1000 mode while deviating for R2000 and R500 modes. This is due to the different nature of the grating efficiency curves provided by the manufacturer (see Table~\ref{table-gratings}). The efficiency data used for R1000 grating is absolute for this grating in littrow configuration. This has a peak efficiency of $\sim$62$\%$ at 5500$\AA$. However, for R2000 and R500 grating, only relative values are given. Due to which their peak efficiencies are higher ($\sim$87$\%$ and $\sim$80$\%$ at 6600$\AA$ and 6800$\AA$ respectively). As these grating efficiency curves are used directly from the manufacturer's datasheets, it causes the simulated efficiencies to be higher for R2000 and R500 modes to generate the simulated performance. The efficiencies of R2000 and R500 modes would also be in good agreement with the derived ones if we consider the grating peak efficiencies to be around 60$\%$ similar to R1000 modes.

\par 
As discussed in Section~\ref{sec-BaseDesign}, three modes of spectroscopy in MFOSC-P were incorporated for varying purposes, viz.  (a) to cover a broad spectral range to determine the shape of SED (R500 mode), (b) to study the evolution of emission lines in bluer part of the spectrum covering H-$\alpha$ and H-$\beta$ emission in a single frame (R1000 mode), and (c) to check for the profile variability of emission lines, in particular for the variability of H-$\alpha$ profile (R2000 mode). These aspects have been verified using suitable targets over the {\bf last couple of years} since the commissioning of the instrument. Some of the observations are described below.\\

\par 
{\bf (1.) Spectroscopy Observations of Symbiotics} \\
\par 

The low/medium resolution optical spectroscopy has been of particular interest in the studies of transient events like outbursts of symbiotic stars and novae. The profile evolution of emission lines is a diagnostic tool to trace the surrounding environment of a symbiotic system (e.g., Nova Sco 2015 \citep{Srivastava2015}). The R2000 mode of MFOSC-P would be of particular use for such cases. Two well known symbiotic stars/novae, T CrB and RS Oph \citep{Anupama1999}, have been observed over the period with MFOSC-P during the on-sky characterization phase.
\par
\begin{figure}
	\centering	
	\includegraphics[width=0.98\textwidth]{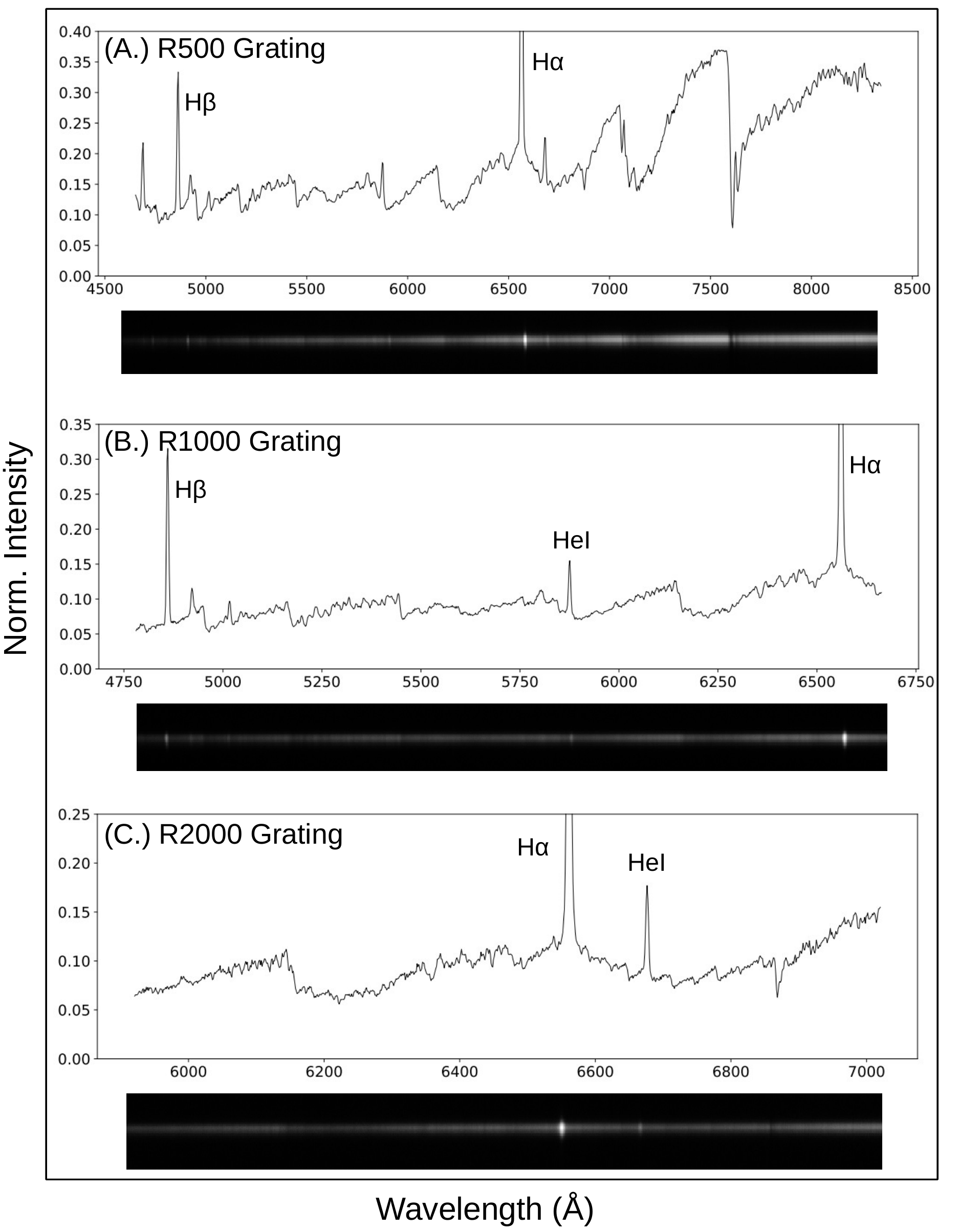}
	\vspace{0.2cm}
	\caption{The spectra of T CrB, a symbiotic system, using all the three spectroscopy modes of MFOSC-P.}
	\label{fig-TCrb}
\end{figure}

 T CrB was observed between 2019 and 2020. Figure~\ref{fig-TCrb} shows the MFOSC-P spectra of T CrB in all the three gratings modes recorded in March 2020. The spectra for 2019 and 2020 appear similar and show the same features as observed during its 2015 super-active state \citep{Munari2016}. Strong emission lines of high ionization species are seen on the top of a nebular continuum with several absorption features of the M giant spectra. MFOSC-P spectra of T CrB for 2019 and 2020 also show several emission lines like He II 4686$\AA$, H-$\beta$, He I 4922$\AA$, He I 5016$\AA$ (blended with [O III] 5007$\AA$), He I 5876$\AA$, H-$\alpha$, He I 6678$\AA$, and He I 7065$\AA$, etc. FWHMs of H-$\alpha$ emissions are measured using R2000 spectra and found to be 7.1$\AA$ and 5.3$\AA$ for 2019 and 2020, respectively, including the instrumental profile. The Equivalent widths (EW) of H-$\alpha$ and H-$\beta$ were estimated after applying the reddening corrections of E$_{B-V}$=0.048 \citep{Munari2016} to R1000 spectra. The EWs for (H-$\alpha$,H-$\beta$) are determined to be (-36.5$\AA$,-19.0 $\AA$) for 2019 and (-49.6$\AA$, -28.1$\AA$) for 2020. These measurements for T CrB are similar to those that were earlier reported by \cite{Munari2016} and \cite{Ilkiewicz2016}.
\par
\begin{figure}
	\centering	
	\includegraphics[width=0.98\textwidth]{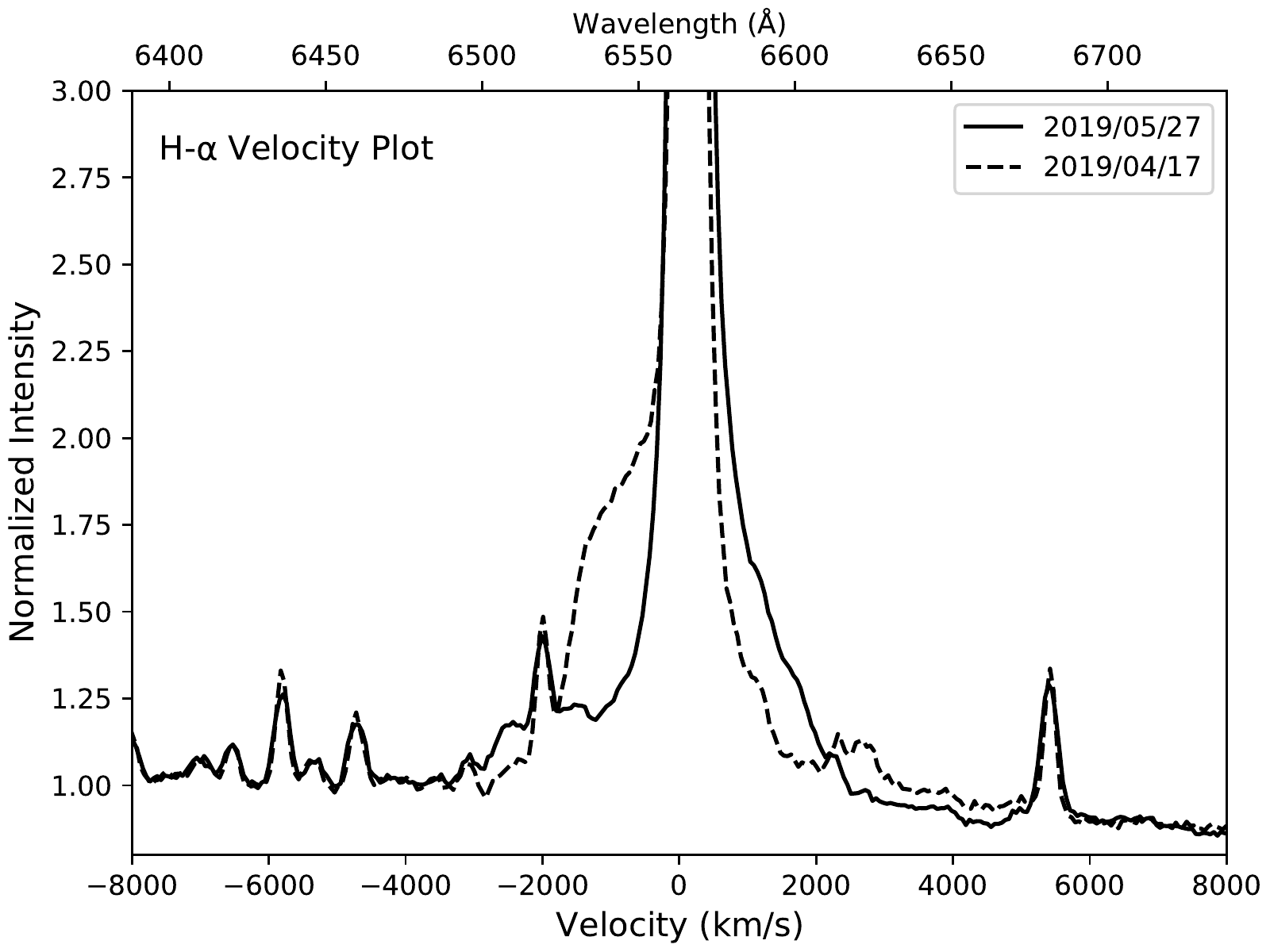}
	\vspace{0.2cm}
	\caption{H-$\alpha$ variability observed in the spectra of RS Oph.}
	\label{fig-RSOph-HalphaProfile}
\end{figure}

We have monitored the spectral evolution of RS Oph for few months. H-$\alpha$ emission in RS Oph is known to show line profile variability over the time scale of a month \citep{Zamanov2005}. MFOSC-P spectra of Rs Oph also show H-$\alpha$ profile variability between April and May 2019. Figure~\ref{fig-RSOph-HalphaProfile} shows line profile variation seen in RS-Oph using the R2000 mode of MFOSC-P. The profile shape and its variation are found to be consistent with the spectra of RS Oph available in the public domain, e.g., ARAS (Astronomical Ring for Access to Spectroscopy\footnote{http://www.astrosurf.com/aras/; Accessed: 2020-04-12}).\\


\begin{figure}
	\centering
	\includegraphics[width=0.98\textwidth]{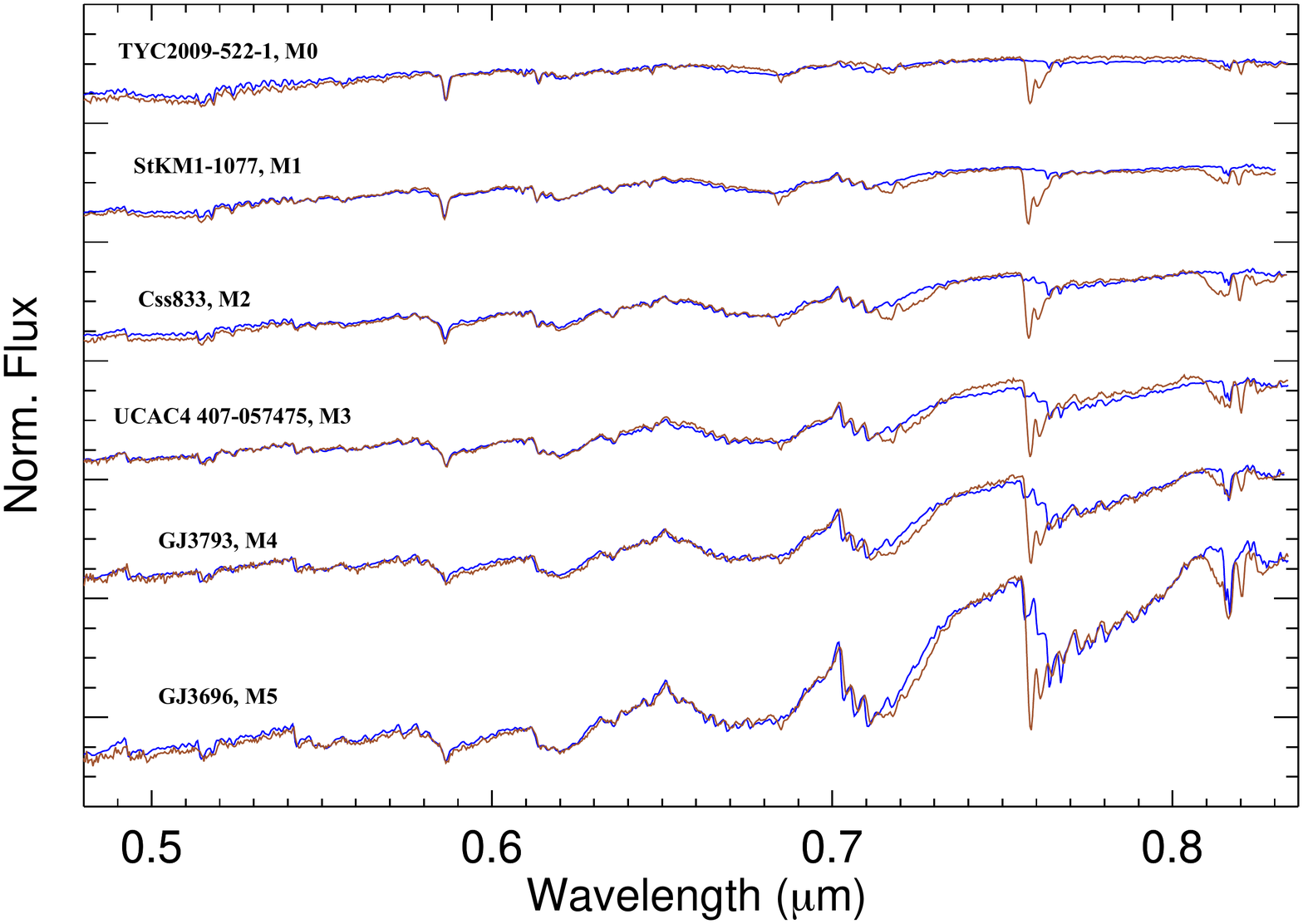}
	\vspace{0.2cm}
	\caption{SDSS template spectra (blue) from \cite{Bochanski2007} is compared with observed spectral sequence of M dwarf using MFOSC-P (red).}
	
	\label{fig-MDwarfs}
\end{figure}


\par 
{\bf (2.) Spectroscopy Observations of M-dwarfs}\\
\par 
Along with the commissioning and characterization run of MFOSC-P, a science program for the characterization of nearby M-dwarfs was undertaken during February-June 2019. Bright M-dwarfs within 100pc with V magnitude less than 14 were selected from the all-sky catalogue of bright M-dwarfs \citep{Lepine2011}. Selected sources cover the spectral types M0 to M5, which is derived from the photometry. A total of 80 targets were observed for their low-resolution spectroscopy using the R500 mode of MFOSC-P. 
\par 
The spectra of these M dwarfs were compared with the SDSS standard M-dwarf template spectra for their sub-classification. Figure~\ref{fig-MDwarfs} shows such a comparison for a sample of M-dwarf spectra. The derived sub-spectral types of M-Dwarfs in our sample ranges from M0-M5 with an error of one spectral subclass. Later the spectral synthesis was performed on these spectra to determine their fundamental stellar parameters, viz. effective temperature and surface gravity. These stellar parameters of our sample targets were determined by comparing the observed spectra with the synthetic spectra generated by the BT-Settl model \citep{Allard2013}. The derived values of effective temperature and surface gravity were in the range from 4000 K to 3000 K and 4.5 to 5.5 dex, respectively. This work has been published as a first science result from MFOSC-P \citep{Rajpurohit2020}. \\

\begin{figure*}
	\centering	
	\includegraphics[width=0.98\textwidth]{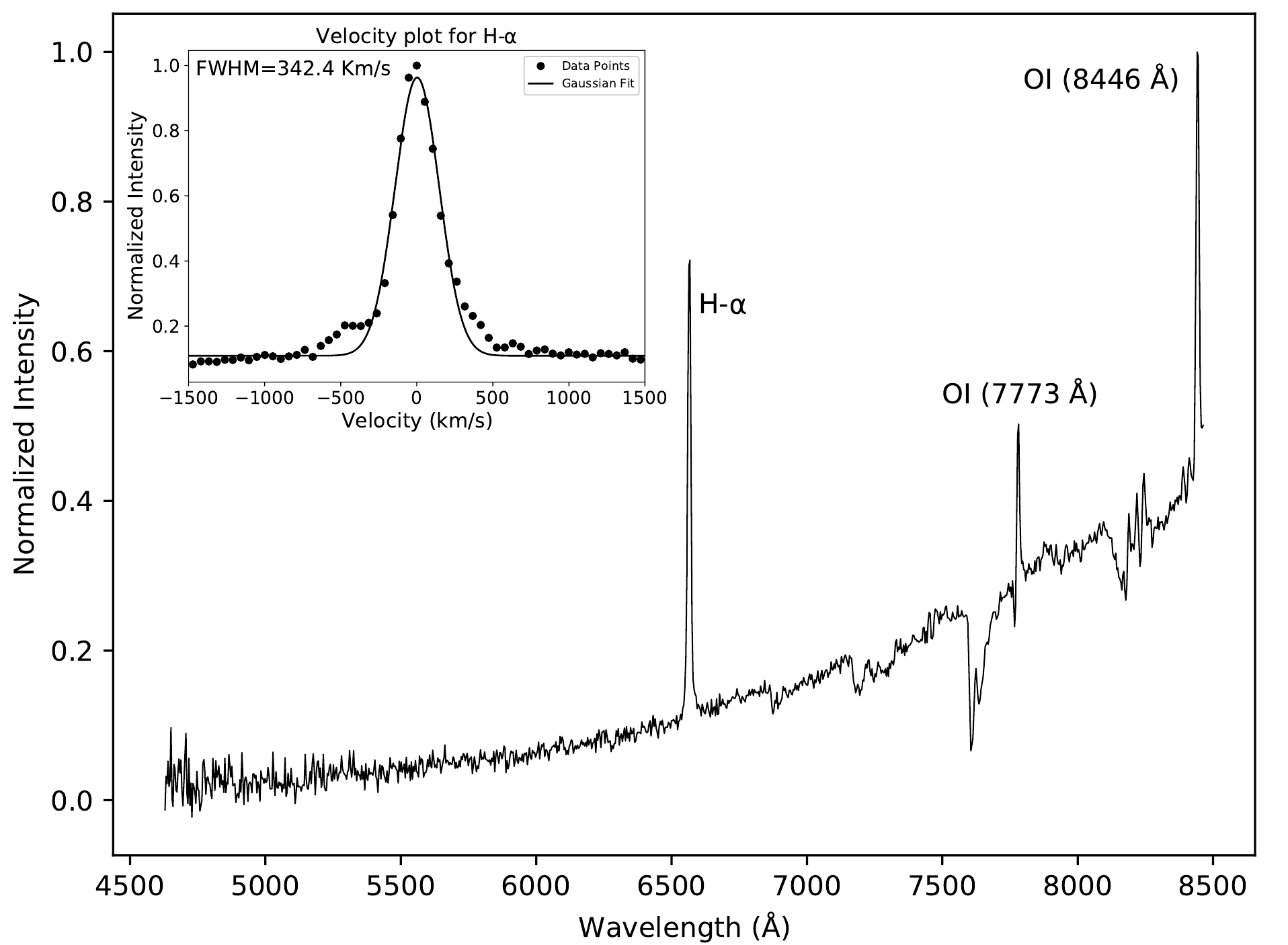}
	\vspace{0.2cm}
	\caption{Low resolution spectrum of Nova AT2019QWF as recorded with MFOSC-P. R500 grating mode was used with integration time of 300 seconds. The nova was $\sim$16 magnitude in V band. Inset: Velocity plot showing H-$\alpha$ profile of Nova AT2019QWF. This was observed with R2000 mode of MFOSC-P.}
	\label{fig-AT2019QWF}
\end{figure*}

\par
{\bf (3.) Spectroscopy of Nova AT2019QWF}\\ 
\par 
Nova AT2019qwf was discovered by \cite{De2019} in a regular survey operation of Palomar Gattini-IR \citep{Moore2019} on UT 2019-09-17.25. The nova was found to be highly reddened, and early spectroscopy \citep{Lee2019} determined the FWHM of the H-$\alpha$ line to be 840 $kms^{-1}$. Optical spectroscopy of this nova using MFOSC-P was first done on UT 2019-11-01.71. The spectra were recorded in the R500 and R2000 modes. The low-resolution spectrum showed a highly reddened continuum with prominent H-$\alpha$ emission and OI lines at 7773$\AA$ and 8446$\AA$. The OI 7773$\AA$ line showed a P-Cygni profile with $\sim$14$\AA$ separation between emission maximum and absorption minimum. FWHM and FWZI (full width at zero intensity) of the H-$\alpha$ line were determined to be $\sim$340 $kms^{-1}$ and 2000 $kms^{-1}$, respectively from the spectrum obtained using the R2000 mode of MFOSC-P. Using the instrumental FWHM of 150 $kms^{-1}$ and quadrature summation method, the intrinsic linewidth was found to be $\sim$300 $kms^{-1}$. Figure~\ref{fig-AT2019QWF} shows the spectrum of the nova on UT 2019-11-01.71 obtained with the R500 mode of MFOSC-P with 300 seconds of integration time.  The velocity plot for H-$\alpha$ (as shown in the inset) was derived from the spectrum obtained using R2000 mode. It is to be noted that nova was of $\sim$16 magnitude in V band on the day of observations, as determined from the AAVSO light curve (AAVSO - American Association of Variable Star Observers\footnote{https://www.aavso.org/; Accessed: 2020-03-20}. The above has been reported to the Astronomical Telegrams \citep{Srivastava2019}.  The nova observations for its spectroscopy evolution continued using MFOSC-P till it went in conjunction with Sun in January 2020. Later, we also obtained the spectra in May 2020 as it came out of the conjunction.

\par
\section{Summary} 
\label{sec-Summary}
\par

This paper presents the design and development details of MFOSC-P on the PRL 1.2m optical-NIR telescope at Mt. Abu. MFOSC-P has been developed as a pathfinder instrument with an aim to develop the next generation FOSC type of instrument for the upcoming PRL 2.5m telescope. The value of MFOSC-P lies in its approach to have a simple, cost-effective optical design and development methodology. Commercially available solutions were utilized in the development of MFOSC-P, wherever possible. The main optical chain is designed as per the requirements of the PRL 1.2m telescope. The optics has been designed with readily available optical glass types and typical industry-standard tolerance values. The overall opto-mechanical system, including the lens barrels, was also designed with moderate tolerance ranges. The lenses of MFOSC-P optical chain were fabricated by a commercial lens manufacturer, whereas the mechanical parts were fabricated using in-house facilities. Other components like filters, grating, slits, CCD detector system, etc. were chosen from commercially available off-the-shelf solutions. Similarly, for the electronics control system and software development, we relied on commercially available products. This approach helped in shortening the project timeline, development efforts, and in cost reduction. 
\par 
The instrument was integrated and assembled in the laboratory using a simple laser retro-reflection method. It was characterized in the laboratory on a test bench set-up using various pinholes and test targets. MFOSC-P was commissioned on the PRL 1.2m telescope in February 2019. A series of commissioning and science observations were carried out from February 2019 till June 2019. The instrument has been used for a variety of science observations since then, e.g., spectroscopy of M-dwarfs, novae, symbiotic systems, H-$\alpha$ detection in star-forming regions, etc.
\par 
MFOSC-P has provided much-needed spectroscopy and multi-filter imaging capability in the visible bands to the PRL 1.2m telescope. As PRL is gearing up towards establishing another 2.5m optical-NIR telescope at Mt. Abu, lessons learned from MFOSC-P are most valuable for the development of the next FOSC instrument for the 2.5m telescope. The design and development approach adopted here is well suited for any small aperture telescope/observatory, which often requires a simple yet versatile instrument within a short period of development. As the instrument has been successfully used for a range of astronomical observations, the MFOSC-P presents a successful FOSC design, which could be adapted with minimal necessary modifications on the similar telescopes around the world.

\begin{acknowledgements}
Development of the MFOSC-P instrument has been funded by the Department of Space, Government of India through Physical Research Laboratory (PRL), Ahmedabad. MFOSC-P team is thankful to the Director, PRL, for supporting the MFOSC-P development program.  We express deep thanks to D.P.K. Banerjee (PRL) and Lalita Sharma (Indian Institute of Technology, Roorkee, India) for useful comments on the manuscript. MKS expresses sincere thanks to Shyam N. Tandon (Inter-University Center for Astronomy and Astrophysics – IUCAA, Pune, India) for detailed discussions on several aspects of FOSC design throughout the development process. VK is thankful to PRL for his Ph.D. research fellowship. MFOSC-P team expresses sincere thanks to Mt. Abu observatory staff for their sustained help and support during MFOSC-P commissioning and subsequent observations. We thank the anonymous reviewer for useful comments and suggestions.		
\end{acknowledgements}

%
\section*{Conflict of interest}
The authors declare that they have no conflict of interest.
\section*{Funding}
The development of the MFOSC-P instrument has been funded by the Department of Space, Government of India through Physical Research Laboratory (PRL), Ahmedabad.
\section*{Availability of data and material}
The instrument characterization data presented in this paper may be made available on reasonable request.
\section*{Code availability}
Not applicable.

\bibliographystyle{spbasic}      

\bibliography{BibList}   

\end{document}